\documentclass[11pt,oneside, a4paper]{article}
\pdfoutput=1

\ifx\pdfoutput\undefined
\usepackage[dvips,bookmarks=false]{hyperref}	% This is for arXiv.org
\else
\usepackage{hyperref}	% This is for pdftex
\fi
\hypersetup{colorlinks,bookmarksopen,bookmarksnumbered,citecolor=blue,
linkcolor=black,pdfstartview=FitH,urlcolor=blue}

\newcommand{\1}{\mbox{1}\hspace{-0.25em}\mbox{l}}

\usepackage{graphicx}
\usepackage{amssymb}
\usepackage{cite}
\usepackage{bm}
\usepackage{indentfirst}
\usepackage{amsmath}
\usepackage{hhline}
\usepackage{multirow}

\def\be{\begin{equation}}
\def\ee{\end{equation}}
\newcommand{\bea}{\begin{eqnarray}}
\newcommand{\eea}{\end{eqnarray}}

\begin{document}
\begin{titlepage}

\begin{flushright}
LPT-ORSAY-15-73
\end{flushright}

\begin{center}

\vspace{1cm}
{\large\bf 
Electric Dipole Moments of Charged Leptons with Sterile Fermions}
\vspace{1cm}

\renewcommand{\thefootnote}{\fnsymbol{footnote}}
Asmaa Abada$^1$\footnote[1]{asmaa.abada@th.u-psud.fr}
and 
Takashi Toma$^1$\footnote[2]{takashi.toma@th.u-psud.fr}
\vspace{5mm}

{\it%{
$^{1}$Laboratoire de Physique Th\'eorique, CNRS, \\
Univ. Paris-Sud, Universit\'e Paris-Saclay, 91405 Orsay, France
\vspace*{.2cm} 
}

\vspace{8mm}

\abstract{
%%%%%%%%%%%%%%%%%%%%%%%%%%%%%%%%%%%%%%%%
We address the impact of sterile fermions on charged lepton electric
 dipole moments. 
Any experimental signal of these observables calls for scenarios of
physics beyond the Standard Model  providing new sources of CP violation. 
In this work,  we consider a minimal extension of the Standard Model 
via the addition of sterile fermions which mix with active neutrinos and we derive  the corresponding analytical expressions for the
 electric dipole moments of charged leptons at two-loop order. 
Our study reveals that, in order to have a non-vanishing contribution in
 this framework, the minimal extension  necessitates the addition of at
 least  2 sterile fermion states to the Standard Model field content. 
Our conclusion is that sterile neutrinos 
can  give significant contributions to the charged lepton electric
 dipole moments, some of them lying within present and future 
experimental sensitivity if the masses of the non-degenerate sterile
 states are both above the electroweak scale. 
The Majorana nature of  neutrinos is also important in order to allow
 for significative contributions to the charged lepton electric dipole
 moments. 
In our analysis we impose all available experimental and observational
 constraints on sterile neutrinos and we further discuss the prospect of
 probing this scenario at low and high energy experiments.
 }

\end{center}
\end{titlepage}

\renewcommand{\thefootnote}{\arabic{footnote}}
\setcounter{footnote}{0}

\setcounter{page}{2}

%%%%%%%%%%%%%%%%%%%%%%%%%%%%%%%%%%%%%%
\section{Introduction}
After the discovery of the Higgs boson at the LHC~\cite{Aad:2012tfa,
Chatrchyan:2012xdj}, the quest of new physics beyond the Standard Model
(SM) is being pursued in many  fronts. Indeed, other than neutrino
oscillation phenomena, there are several observational problems and
theoretical caveats  suggesting that 
new physics is indeed required: the former are related to the baryon asymmetry of
the Universe and the need for a 
dark matter candidate, while among the latter one can mention the
hierarchy problem, the flavour puzzle, or fine-tuning  
in relation to electroweak (EW) symmetry breaking. Disentangling the new
physics models and in particular, probing the underlying neutrino mass generation mechanism, 
requires investigating all available observables, arising from all
avenues - high-intensity, high-energy and 
cosmology - as well as thoroughly exploring the interplay between them. 

There are numerous well-motivated and appealing models of new physics that aim at
addressing these issues, and which are currently being actively
investigated and searched for. 
Focusing on the neutrino oscillation phenomena, in order to account for
neutrino masses and mixings, many extensions of the SM
call upon the introduction of right-handed neutrinos - which are
sterile states - giving rise to a Dirac mass term for the neutral leptons.  
One of the most economical possibility is the embedding of the (standard
or type I) seesaw
mechanism~\cite{Minkowski:1977sc,Yanagida:1979as,GellMann:1980vs,Glashow:1979nm,Mohapatra:1979ia, 
Schechter:1980gr,Schechter:1981cv} into the SM. 
These scenarios cannot be probed directly: in
order to have natural Yukawa couplings  for the neutral leptons, the mass 
of the right-handed neutrinos must be in general very high, 
close to the gauge coupling unification scale, thus implying that direct experimental tests of
the seesaw hypothesis are virtually impossible.  
In contrast, low-scale seesaw models~\cite{Mohapatra:1986bd,GonzalezGarcia:1988rw,Deppisch:2004fa,Asaka:2005an,Gavela:2009cd,Ibarra:2010xw,Abada:2014vea}, in which right-handed or sterile fermion states with masses
around the electroweak scale or even much lower are added to the SM,
prove to be very attractive since the new states can be 
produced in colliders and/or in  low-energy experiments, and their 
contributions to physical processes can be sizable, provided that a
non-negligible mixing to the  (mostly) active neutrinos is present. 
This is the case 
for the $\nu$MSM~\cite{Asaka:2005an}, the Inverse
Seesaw~\cite{Mohapatra:1986bd}, the Linear
Seesaw~\cite{Barr:2003nn,Malinsky:2005bi} and the low-scale 
type-I seesaw~\cite{Gavela:2009cd,Ibarra:2010xw}. 

Some of the latter  models may also explain the relic density of dark matter
taking the lightest keV scale sterile neutrino as a candidate~\cite{Asaka:2005an,Abada:2014zra} 
and the baryon asymmetry of the Universe obtained via leptogenesis through neutrino
oscillations and CP violating phases in the lepton
sector~\cite{Akhmedov:1998qx,Canetti:2012vf,Canetti:2012kh,Abada:2015rta}. 
In addition, low scale (GeV-scale) leptogenesis with $3$ sterile neutrinos has also been
discussed~\cite{Canetti:2014dka}.

Present in several neutrino mass models, the sterile fermion masses can range from
well below the electroweak scale (GeV) up to the Planck scale. 
Their existence  is strongly motivated
by current  reactor~\cite{Mueller:2011nm, Huber:2011wv, Mention:2011rk},
accelerator~\cite{Aguilar:2001ty,AguilarArevalo:2007it,AguilarArevalo:2010wv,Aguilar-Arevalo:2013pmq}  and Gallium
anomalies~\cite{Acero:2007su,Giunti:2010zu} suggesting that there might be some extra
fermionic gauge singlets with mass(es) in the eV range. 
Other than the reactor and accelerator anomalies, their existence  is
also motivated  by  indications from large scale structure
formation~\cite{Kusenko:2009up,Abazajian:2012ys}. Moreover and depending
on their masses, sterile fermion  states present in these extensions can
also give rise to interesting collider
signatures~\cite{Atre:2009rg,Dev:2013wba,Adams:2013qkq,Bonivento:2013jag,Blondel:2013ia,Deppisch:2015qwa,Alekhin:2015byh,Anelli:2015pba,Banerjee:2015gca,deGouvea:2015euy,Das:2014jxa,Das:2015toa,Das:2012ze}. Nevertheless,
due to the mixings of the sterile  states with the active left-handed
neutrinos,  models with sterile fermions are severely constrained from
electroweak precision observables, laboratory data and cosmology. 

These extensions of the SM with sterile fermions aiming at incorporating
massive neutrinos and the leptonic mixing may also open the door  
to many new phenomena, such as flavour violation in the charged lepton
sector and contributions to lepton flavour-conserving
observables such as charged lepton electric and magnetic dipole 
moments.

In this work,  we study the effect of sterile fermions which mix with
 active neutrinos  on the electric dipole moments (EDMs) of charged
 leptons. 
 An EDM, which corresponds to the coupling of the spin of a particle to
 an external electric field,  is a flavour conserving  observable which
 may provide a clear signal of the existence of extra CP violating
 phases. 

Exploring the origin of CP violation is important in extending the SM. 
The CP violating observables which have been identified in  Kaon and B meson systems
are  consistent with the SM explanations and their measured values can
be accommodated with the unique source of CP violation of the
SM.\footnote{There is an additional CP violating term 
$\mathcal{L}=\theta_\mathrm{QCD}\frac{\alpha_s}{8\pi}G_{\mu\nu}^a\tilde{G}^{a\mu\nu}$, where
$\tilde{G}^{a\mu\nu}\equiv\frac{1}{2}\epsilon^{\mu\nu\rho\sigma}G_{\rho\sigma}^a$
is the dual tensor of the field strength tensor of gluon
$G_{\mu\nu}^a$ and $\alpha_s$ is the strong coupling constant. This coupling ($\theta_\mathrm{QCD}$) should be 
$\theta_\mathrm{QCD}\lesssim10^{-10}$ due to the constraint of the neutron EDM.}
The amount of CP violation in the SM can be parametrised by  the Jarlskog invariant
$J_{CP}$.
However, in order to explain the observed baryon asymmetry of the
Universe, additional CP-violating 
sources are required. Extensions of the SM accommodating 
 neutrino masses and mixings might provide additional sources of CP
violation. In fact, by itself, the parametrisation of the leptonic
mixing matrix in  terms of 3 mixing angles and one Dirac CP violating
phase (plus two additional ones in the case where the neutrinos are of
Majorana nature) already opens the possibility of  CP violation in the
leptonic sector.

For this study, we consider a minimal extension of the SM 
via the addition of $N$ of sterile fermions which mix with the 
active neutrinos, and we address their impact on the charged lepton
EDMs. In our chosen framework, we do not impose any seesaw realization,
meaning  no hypothesis is made on the underlying mechanism of neutrino masses
and mixings generation, only assuming that the physical and the interaction
neutrino basis are related via a $(3+N) \times (3+N)$ unitary mixing
matrix,  which reduces to the Pontcorvo-Maki-Nakagawa-Sakata (PMNS) matrix, $U_\text{PMNS}$, in the case
of  three neutrino generations (no additional sterile neutrinos are
present). 
Since we consider that neutrino mass eigenvalues and the lepton mixing
matrix are independent, this simple model can be seen as an "effective"
model allowing  to investigate  the potential imprints (on  EDMs) due to
the presence of a number  $N$ of sterile neutrinos present in several new physics
scenario including Type-I seesaw, inverse seesaw and linear seesaw
models. 
The computation of EDMs in the presence of right-handed neutrinos has been
extensively discussed in
Refs.~\cite{Ng:1995cs,Archambault:2004td,Chang:2004pba}, and that in
the context of supersymmetric seesaw has been discussed in
Ref.~\cite{Ellis:2001yza, Masina:2003wt, Farzan:2004qu}. 

In the presence of massive sterile neutrinos, charged lepton EDMs are induced at
two-loop level. We have computed the corresponding diagrams -
providing the corresponding analytical expressions for the EDMs
of charged leptons at two-loop order - and we have shown that  in order
to have a significant contribution, the minimal extension of the
SM indeed requires the addition of at least  2 sterile
fermion states.  We have also shown that the Majorana nature of
neutrinos is also important in order to allow  for significative
contribution to the charged lepton EDMs. 
We complete our analysis by  also discussing  the several
experimental and theoretical constraints on our scenario including those
from  charged lepton flavour violating (cLFV) processes, direct collider searches, electroweak 
precision data and the perturbative unitarity constraint. 
We confront our findings to the current experimental status and we
conclude that, depending on their masses  and on the 
active-sterile mixing angles,  sterile neutrinos 
can give significant contributions to the charged lepton EDMs, some of
these observables even  lying within present and future 
experimental sensitivity.

%%%%%%%%%%%%%%%%%%%%%%%%%%%%%%%%%%%%%%
\section{The Model}\label{sec:model}
In order to accommodate neutrino masses and
mixings, the SM can be extended with new  sterile fermions 
such as right-handed (Majorana) neutrinos.
In this work, we consider the SM extended by 
$N$ sterile fermion states which mix with the three active neutrinos. 
We consider that the neutrino mass eigenvalues and the lepton mixing
matrix are independent, meaning that no assumption is made on the
neutrino mass generation mechanism.\footnote{Should we have considered a given neutrino mass generation mechanism, the physical  neutrino masses and the lepton mixing matrix $U$ would be derived 
from the diagonalisation of the full  $(3+N)\times (3+N)$ neutrino mass  matrix and thus be related - provided one respects neutrino data in what concerns the active (light) neutrinos.} 
As we will see later, we focus on the $3+1~(N=1)$ and $3+2~(N=2)$
models. For $N>2$, we expect that the results do not change 
with respect to the $N=2$ case.

\subsection{Lagrangian}\label{sec:lag}
After electroweak symmetry breaking, the relevant terms in the
Lagrangian can be written in the Feynman-'t Hooft gauge as
\begin{eqnarray}
\mathcal{L}\hspace{-0.2cm}&=&\hspace{-0.2cm}
-\frac{g_2}{\sqrt{2}}U_{\alpha
 i}W_\mu^{-}\overline{\ell_\alpha}\gamma^{\mu}P_L\nu_i
-\frac{g_2}{\sqrt{2}}U_{\alpha
i}H^{-}\overline{\ell_\alpha}\left(\frac{m_\alpha}{m_W}P_L-\frac{m_i}{m_W}P_R\right)\nu_i 
+\mathrm{H.c.}\nonumber\\
&&\hspace{-0.2cm}
-\frac{g_2}{2\cos\theta_W}U_{\alpha i}^*U_{\alpha
j}Z_{\mu}\overline{\nu_i}\gamma^{\mu}P_L\nu_j
-\frac{ig_2}{2}U_{\alpha i}^*U_{\alpha j}A^0\overline{\nu_i}\left(\frac{m_j}{m_W}P_R\right)\nu_j
+\mathrm{H.c.}\nonumber\\
&&\hspace{-0.2cm}
-\frac{g_2}{2}U_{\alpha i}^*U_{\alpha
j}h\overline{\nu_i}\left(\frac{m_j}{m_W}P_R\right)\nu_j
+\mathrm{H.c.}\ , 
\label{eq:lag}
\end{eqnarray}
where $g_2$ is the $SU(2)_L$ gauge coupling, $U_{\alpha i}$ is the
$3\times(3+N)$ lepton mixing matrix, 
$m_i$ is the mass eigenvalue of the neutrinos and $m_\alpha$ is the charged lepton mass. 
The indices $\alpha$ and $i,j$ run $\alpha=e,\mu,\tau$ and $i,j=1,\cdots,3+N$.
The mixing matrix $U_{\alpha i}$ obeys the following relations due to unitarity conditions:
\begin{equation}
\sum_{i=1}^{3+N}U_{\alpha i} U^*_{\beta i} =\delta_{\alpha\beta},\quad
\sum_{\alpha=e,\mu,\tau}U^*_{\alpha i}
U_{\alpha j} \neq\delta_{ij},
\end{equation}
Further details can be found in, for example, Refs.~\cite{Ilakovac:1994kj,Alonso:2012ji}. 
Although the above Lagrangian in Eq.~(\ref{eq:lag}) has been derived by
assuming Type-I seesaw mechanism in Ref~\cite{Alonso:2012ji}, it is 
also valid for Inverse seesaw and Linear seesaw
mechanisms as well. The difference among these mechanisms is reflected in 
the mixing matrix $U_{\alpha i}$ obtained after the diagonalisation of the corresponding neutrino mass
matrix. 
However Eq.~(\ref{eq:lag}) is not valid for a (pure) Type-II seesaw mechanism due to the presence of 
$SU(2)_L$ scalar triplet(s) instead of sterile fermions. 
%with hypercharge $Y=1$ is introduced instead of sterile fermions. 
%In the case of a  Type-III seesaw mechanism where  $SU(2)_L$ fermion
%triplet(s) (with, for instance, hypercharge $Y=0$) is (are) present(s), 
%the charged components of fermion triplet mix with the charged leptons of 
%the SM. 
%Moreover the neutral component of the triplet is not sterile since it has the gauge interactions. 
%Thus the Lagrangian in Eq.~(\ref{eq:lag}) is slightly extended for Type-III
%seesaw mechanism such that the index $\alpha=e,\mu,\tau$ to
%$\alpha^{\prime}=e,\mu,\tau,1,\cdots, N^{\prime}$ 
%where $N^{\prime}$ is the number of the triplet fermion. 

\subsection{Mixing matrix}
In the $3+N$ model, the mixing matrix $U$ includes $(3+N)(2+N)/2$ rotation angles,
$(2+N)(1+N)/2$ Dirac phases and $2+N$ Majorana phases.
As an example, the mixing matrix $U$ for the $N=2$ can be parametrised as 
\begin{equation}
 U=R_{45}R_{35}R_{25}R_{15}R_{34}R_{24}R_{14}R_{23}R_{13}R_{12}
~\mathrm{diag}\left(1,e^{i\varphi_2},e^{i\varphi_3},e^{i\varphi_4},e^{i\varphi_5}\right),
\label{eq:mixing}
\end{equation} 
where $R_{ij}$ is the rotation matrix between $i$ and $j$. 
For instance, the rotation matrix $R_{45}$ is explicitly given by 
\begin{equation}
R_{45}=\left(
\begin{array}{ccccc}
1 & 0 & 0 & 0 & 0\\
0 & 1 & 0 & 0 & 0\\
0 & 0 & 1 & 0 & 0\\
0 & 0 & 0 & \cos\theta_{45} & \sin\theta_{45}e^{-i\delta_{45}}\\
0 & 0 & 0 & -\sin\theta_{45}e^{i\delta_{45}} & \cos\theta_{45}\\
\end{array}
\right),
\end{equation}
and likewise for the other matrices $R_{ij}$ (in terms of $\theta_{ij}$ and  $\delta_{ij}$).

Since the number of Dirac phases is 6 for the case where $N=2$, four Dirac phases
$\delta_{ij}$ can be eliminated. 
In this paper, we set $\delta_{12}=\delta_{23}=\delta_{24}=\delta_{45}=0$. 
The mixing matrix for $N=1$ can be obtained by taking the $4\times4$
sub-matrix after substituting $R_{i5}=\1$ in Eq.~(\ref{eq:mixing}).

According to a global analysis of solar, atmospheric, reactor and accelerator
neutrino data, the best fit values of the mixing angles and neutrino
mass differences for normal hierarchy are given by~\cite{Gonzalez-Garcia:2014bfa}, 
\begin{eqnarray}
&&
\sin^2\theta_{12}=0.304,\quad
\sin^2\theta_{23}=0.452,\quad
\sin^2\theta_{13}=0.0218,\quad\nonumber\\
&&
\Delta{m}_{21}^2=7.50\times10^{-5}~\mathrm{eV},\quad
\Delta{m}_{31}^2=2.457\times10^{-3}~\mathrm{eV},\nonumber
\end{eqnarray}
and for inverted hierarchy are given by
\begin{eqnarray}
&&
\sin^2\theta_{12}=0.304,\quad
\sin^2\theta_{23}=0.579,\quad
\sin^2\theta_{13}=0.0219,\quad\nonumber\\
&&
\Delta{m}_{21}^2=7.50\times10^{-5}~\mathrm{eV},\quad
\Delta{m}_{31}^2=-2.449\times10^{-3}~\mathrm{eV}.\nonumber
\end{eqnarray}
We use these values for our numerical analyses.
Note that the following EDM computation does not substantially depend on
neutrino mass hierarchies since, as we will see, the heavy sterile neutrinos give the dominant
contributions to the EDMs.

In our work we have varied all the phases\footnote{The fitting of the CP violating Dirac phase
$\delta_{13}$ also has been done in Ref.~\cite{Gonzalez-Garcia:2014bfa}, and
the Dirac phase is in the range of $236^\circ<\delta_{13}<345^\circ$ at
$1\sigma$ level, but all values are allowed at $3\sigma$ level.}  including $\delta_{13}$ in the range 
$[0, 2\pi]$.

%%%%%%%%%%%%%%%%%%%%%%%%%%%%%%%%%%%%%%
\section{Electric Dipole Moments}

A non-zero value of the EDM for elementary particles implies violations of parity
(P) and time reversal (T) symmetries. This is
translated to CP violation due to requirement of CPT invariance.
In the SM, the electron EDM is induced at four-loop level through
the Jarlskog invariant $J_{CP}$, and
the predicted value is approximately  given by~\cite{Pospelov:2005pr, Fukuyama:2012np}
\begin{equation}
|d_e|/e\sim
\frac{\alpha_W^3\alpha_sm_e}{256(4\pi)^4m_W^2}J_{CP}\sim
3\times10^{-38}~\mathrm{cm},
\label{eq:edm_sm}
\end{equation}
where the Jarlskog invariant is defined with the CKM matrix $V$ by
\begin{equation}
J_{CP}\equiv|\mathrm{Im}\left(V_{us}V_{cs}^*V_{cb}V_{ub}^*\right)|
=\sin\theta_{12}^q\sin\theta_{23}^q\sin\theta_{13}^q\cos\theta_{12}^q
\cos\theta_{23}^q\cos^2\theta_{13}^q\sin\delta_q, 
\end{equation}
with the rotation angles $\theta_{12}^q$, $\theta_{23}^q$,
$\theta_{13}^q$ and the Dirac CP phase $\delta_q$ in the quark sector. 
The  value of the Jarlskog invariant obtained  from the global fit of the CKM matrix elements 
is  
$J_{CP}=3\times10^{-5}$~\cite{Agashe:2014kda}. 
The electron EDM value in Eq.~(\ref{eq:edm_sm}) is too small
compared to the current experimental limit 
$|d_e|/e<8.7\times10^{-29}~\mathrm{cm}$ set  by ACME Collaboration~\cite{Baron:2013eja}. 

\subsection{Current experimental bounds}
The present experimental upper bounds for the charged lepton EDMs are given by
\begin{eqnarray}
\left|d_e\right|/e\hspace{-0.2cm}&<&\hspace{-0.2cm}
8.7\times10^{-29}~[\mathrm{cm}]\quad(\mathrm{ACME}),\\
\left|d_\mu\right|/e\hspace{-0.2cm}&<&\hspace{-0.2cm}
1.9\times10^{-19}~[\mathrm{cm}]\quad(\mathrm{Muon}~g-2),\\
\left|\mathrm{Re}\left(d_\tau\right)\right|/e\hspace{-0.2cm}&<&\hspace{-0.2cm}
4.5\times10^{-17}~[\mathrm{cm}]\quad(\mathrm{Belle}),\\
\left|\mathrm{Im}\left(d_\tau\right)\right|/e\hspace{-0.2cm}&<&\hspace{-0.2cm}
2.5\times10^{-17}~[\mathrm{cm}]\quad(\mathrm{Belle}).
\end{eqnarray}
The bounds for the electron, the muon and the tau EDMs have been measured by  the ACME
Collaboration~\cite{Baron:2013eja}, 
the Muon $(g-2)$ Collaboration~\cite{Bennett:2008dy} and the Belle
Collaboration~\cite{Inami:2002ah, Agashe:2014kda} respectively. 
The bound for the electron EDM is especially strong. 
EDMs are normally understood as  real numbers since one works under the assumption of
CPT invariance. However, the Belle Collaboration is attempting to
measure an effect of CPT violation as well, which implies violation of Lorentz
invariance~\cite{Greenberg:2002uu} and thus, they also give an upper bound on 
$\mathrm{Im}\left(d_\tau\right)$ as a CPT violating parameter. 

The  next generation ACME experiment is expected to be able to reach a
sensitivity for the electron EDM $\sim\
|d_e|/e\lesssim\mathcal{O}(10^{-30})~\mathrm{cm}$ with more molecules 
and smaller systematics~\cite{acme:next_generation}. 
The future sensitivity for the muon EDM is
$|d_\mu|/e\sim10^{-21}~\mathrm{cm}$ by J-PARC $g-2$/EDM
Collaboration~\cite{Saito:2012zz}.

%%%%%%%%%%%%%%%%
\subsection{Calculation of charged lepton EDMs}
\label{calculation}

In the $3+N$ model, EDMs for the charged leptons are not induced
at one-loop level since the relevant 
amplitude is always proportional to $|U_{\alpha i}|^2$ 
which is a purely real number. Imaginary parts of the amplitude are essential for EDMs.
Thus, the leading contributions to charged lepton EDMs are given at two-loop
level and the relevant diagrams are depicted in Fig.~\ref{fig:fig1}. 
In this work, we compute all the latter contributions in 
the Feynman-'t Hooft gauge meaning that each of the diagrams of
Fig.~\ref{fig:fig1} stands for all possible combinations arising when one
(or more) gauge boson is replaced by its corresponding Goldstone
boson. 
\subsubsection{Diagrams}
\begin{figure}[h!]
\begin{center}
\includegraphics[scale=0.59]{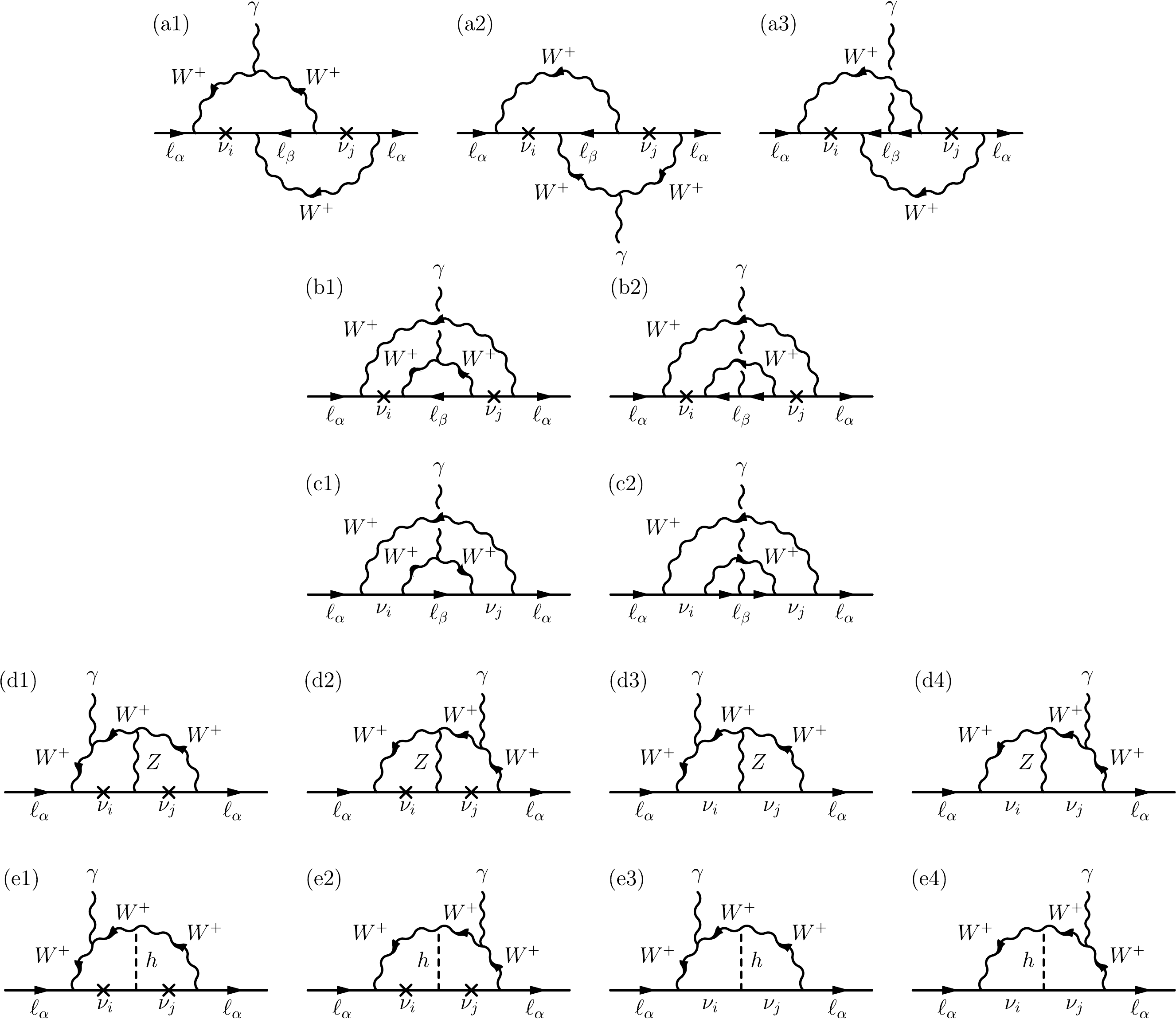}
\caption{Diagrams contributing to charged lepton EDMs at two-loop level. 
The cross in the $\nu_i$, $\nu_j$ propagators means lepton number
 violation due to Majorana neutrino nature.}
\label{fig:fig1}
\end{center}
\end{figure}

\begin{figure}[t]
\begin{center}
\includegraphics[scale=0.59]{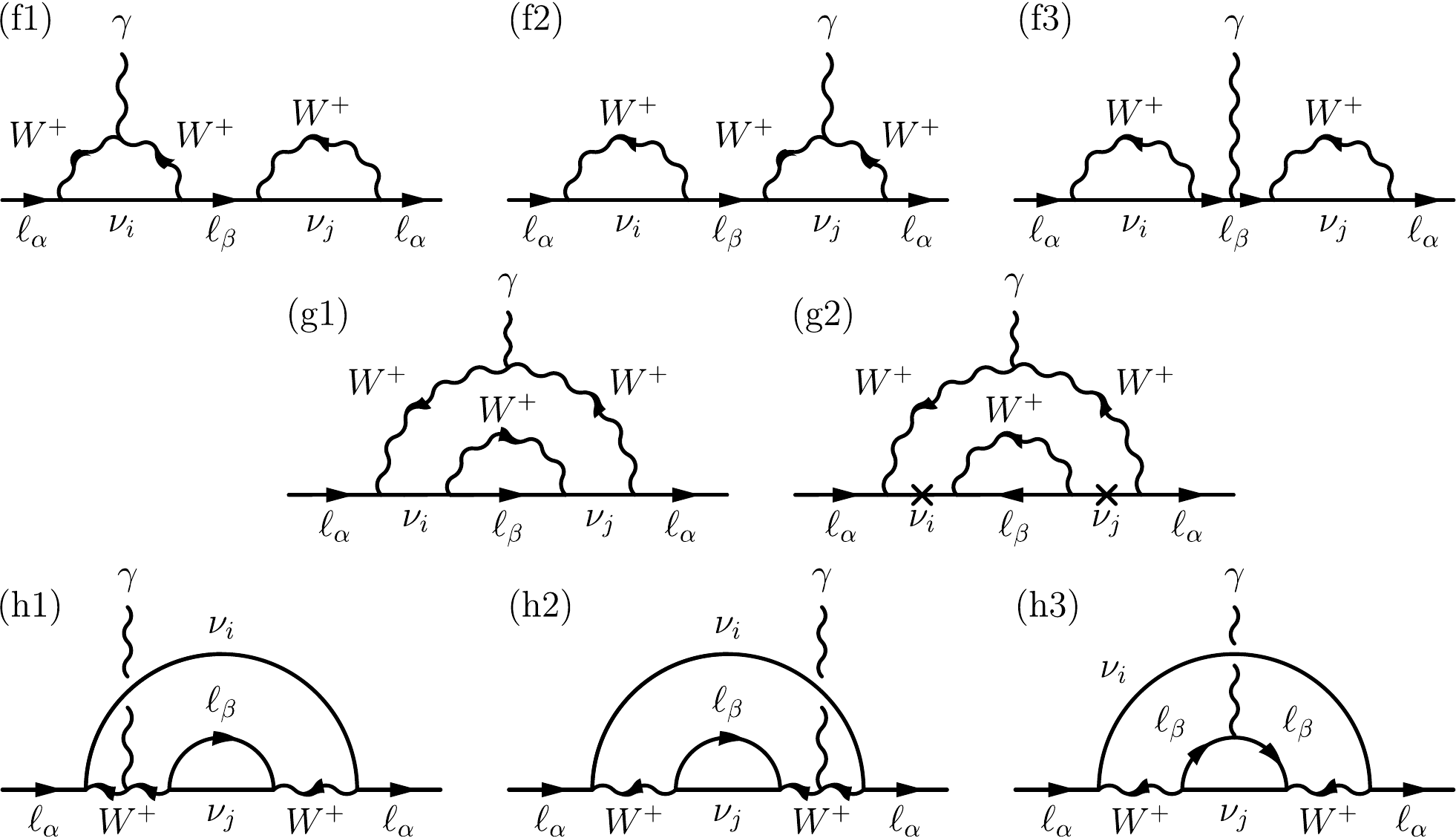}
\caption{Diagrams not contributing to charged lepton EDMs at two-loop level.}
\label{fig:fig2}
\end{center}
\end{figure}

Some comments are in order before proceeding to the computation of the EDMs: 
\begin{itemize}

\item The diagrams (a1), (a2), (a3), (b1) and (b2) exist only if the neutrinos are
      Majorana fermions, and give a non-zero EDM.
      On the other hand  diagrams (c1) and (c2) are non-zero for both cases
      of Majorana and Dirac neutrinos. 
      If, in analogy to what occurs in the quark sector, the model includes 
      a single Dirac CP-violating phase in the lepton sector, 
      the contribution of the diagrams (c1) and (c2) would be zero as
      discussed in detail in Ref.~\cite{Shabalin:1978rs}. 
\item In our computation, the diagrams (d1), (d2), (d3), (d4), (e1) (e2), (e3)
      and (e4),  mediated by the $Z$ and the Higgs bosons (last two
      lines in Fig.~\ref{fig:fig1}) will give similar (same order)
      contributions compared to the ones of the diagrams (a1), (a2),
      (a3), (b1), (b2), (c1) and (c2) - the only difference being a charged lepton
      propagator for the latter diagrams and the extra Higgs or $Z$
      boson propagators for the former ones. 
      Should one include them in the computation, their effect would
      translate into a small factor difference at most and will not
      change qualitatively the results. 
      In addition,  for  diagrams (e)  (the last  line of diagrams of
      Fig. 1), with the Higgs propagating in the loops, 
   the Goldstone modes  give rise to contributions  of the order of the
      ones from diagrams (a), (b) and (c), as well as an
      additional contribution due to the 
      Higgs self-coupling $h|H^+|^2$. 
    The same argument holds for the diagrams with a $Z$ propagating in the loops of 
      diagrams in the fourth line of Fig. 1, corresponding to diagrams (d),  since the
      Weinberg angle is introduced in the corresponding contribution
      which also avoids systematic cancellation. 

\item On the other hand, the diagrams in Fig.~\ref{fig:fig2} do not
      contribute to charged lepton EDMs. This can be understood as follows. 
      The diagrams (f1), (f2) and (f3) in Fig.~\ref{fig:fig2} include the
      self-energy correction of charged leptons at one-loop level. 
      The self-energy is decomposed into two parts, a dispersive 
      and an absorptive one. The resultant dispersive part
      corresponds to a  renormalisation of  
      the wave functions and of the charged lepton masses, whereas the absorptive part gives
      a contribution to physical quantities and exists
      only if the intermediate particles in the loop are
      on-shell.\footnote{Leptogenesis is one of the examples for such a
      case~\cite{Pilaftsis:1997dr, Buchmuller:1997yu}.} 
      Since the charged leptons cannot decay into a gauge boson and the
      neutrino, in our case  we do not have absorptive parts of the
      self-energy. Thus after renormalisation, the self-energy contribution becomes
      diagonal in terms of $\alpha$ and $\beta$, and the diagrams do not give 
      a contribution to the EDMs since an off-diagonal component $(\alpha\neq\beta)$
      is needed in order to have a contribution. 
\item For the diagrams  (g1) and (g2) in Fig.~\ref{fig:fig2}, the
      relevant amplitudes for EDMs are in general factorised into CP
      phase factors and dimensionless loop functions after some
      calculations. 
      As we will see later, the CP phase factors and loop functions
      should be both anti-symmetric under the exchange 
      $i\leftrightarrow j$ in order to give a non-zero contribution to
      the EDM ($i,j$ referring to the neutrino mass eigenstates). 
      However, since the resulting loop functions for the diagrams (g1) and (g2) 
      are fully symmetric under the exchange $i\leftrightarrow j$,
      their contributions to EDMs are thus zero. 
\item The diagrams (h1), (h2) and (h3) in Fig.~\ref{fig:fig2} do not
      contribute to EDMs since, as one can notice, their amplitudes
      are always proportional to $|U_{\alpha i}|^2|U_{\beta j}|^2$,
      which are real quantities. 
\end{itemize}
As a consequence of the above discussion, we are left with 44 two-loop
diagrams for (a1), (a2), (a3), (b1), (b2), (c1) and 
(c2), taking into account longitudinal modes of 
the Goldstone boson. 
(Should we take into account the $Z$ and Higgs bosons contributions,
which correspond to the diagrams (d1), (d2), (d3), (d4), (e1) (e2), (e3)
and (e4) of Fig.~\ref{fig:fig1}, this would correspond to  96  two-loop
diagrams. However, as mentioned above, we do not consider the latter
diagrams (d) and (e).) 
While some additional diagrams given by four-point vertices
$h$-$H^{\pm}$-$W^{\mp}$-$\gamma$, $Z$-$H^{\pm}$-$H^{\mp}$-$\gamma$,
$A^{0}$-$H^{\pm}$-$W^{\mp}$-$\gamma$ $Z$-$W^{\pm}$-$W^{\mp}$-$\gamma$ also exist, 
we have checked that these contributions are negligible compared to the
relevant diagrams we considered. 

\subsubsection{Some relevant steps in the calculation of the EDMs}
\label{steps}

We perform the computation of the charged lepton EDMs as follows. 
After having written the loop integrals with the help of the
Feynman parameters, the simplification of the amplitude 
is done with FeynCalc~\cite{Mertig:1990an}. 
Finally, one can extract the relevant terms to the EDMs, 
\begin{equation}
i\mathcal{M}=d_\alpha\epsilon_{\mu}^*(q)\overline{u}(p_2)i\sigma^{\mu\nu}q_\nu\gamma_5
 u(p_1),
\end{equation}
where $d_\alpha$ is the EDM of a charged lepton $\ell_\alpha$, $p_1$
 and $p_2$ are the initial and final momentum of the charged lepton,
 respectively, and finally, $q\ (q= p_2-p_1)$ is the photon momentum.%

In the $3+N$ model, the EDM of a charged lepton $\ell_\alpha$ can be expressed by
\begin{equation}
d_\alpha=-\frac{g_2^4\ e\ m_\alpha}{4(4\pi)^4m_W^2}\sum_\beta\sum_{i,j}
\biggl[
J_{ij\alpha\beta}^{M}
I_M\left(x_i,x_j,x_\alpha,x_\beta\right) 
+J_{ij\alpha\beta}^{D} I_D\left(x_i,x_j,x_\alpha,x_\beta\right)
\biggr],
\label{eq:edm}
\end{equation}
where 
\begin{equation}
J_{ij\alpha\beta}^M\equiv\mathrm{Im}\left(U_{\alpha j}U_{\beta j}U_{\beta
i}^*U_{\alpha i}^*\right)\ \quad \text{and}\quad 
J_{ij\alpha\beta}^D\equiv\mathrm{Im}\left(U_{\alpha j}U_{\beta j}^{*}U_{\beta
i}U_{\alpha i}^*\right)\ , 
\label{eq:edm:IJ}
\end{equation}
are the phase factors obtained from the
relevant Majorana and Dirac type diagrams. The dimensionless loop
functions $I_M$ and $I_D$ are expressed in terms of 
 the variables $x_A\equiv m_A^2/m_W^2$
($A=i,j,\alpha,\beta$); the loop functions of the dominant part are given in the Appendix. 
The first term in Eq.~(\ref{eq:edm}) comes from the diagrams (a1),
(a2), (a3), (b1) and (b2), while the second term arises  from the (c1) and
(c2) diagrams. 
Notice that the loop function $I_M$ is always proportional to
$m_im_j/m_W^2$ because of the nature of the Majorana neutrinos.
One can also see from the definition in Eq.~(\ref{eq:edm:IJ}) that
the phase factors $J_{ij\alpha\beta}^M$ and 
$J_{ij\alpha\beta}^D$ are anti-symmetric 
under the exchange of $i\leftrightarrow j$. 
As a result, only the anti-symmetric part of the loop functions $I_M$
and $I_D$ is relevant. 

Since the loop function $I_M$ is always proportional to
$m_im_j/m_W^2$, it is convenient to extract this ratio as an overall factor. 
Moreover since the charged lepton mass is much lighter than the $W$
gauge boson mass, we can take $x_\alpha,x_\beta\approx0$ as a good
approximation if $m_i,m_j\gg m_\alpha,m_\beta$ ($i,j\ > 3$ since, 
as one can see below, when the masses $m_i$ and $m_j$ are lighter than
the charged lepton masses, the contributions to EDMs due to such light
sterile states are extremely small). 
For all these reasons,  Eq.~(\ref{eq:edm}) can be simplified to 
\begin{eqnarray}
d_\alpha\hspace{-0.2cm}&=&\hspace{-0.2cm}
-\frac{g_2^4\ e\ m_\alpha}{4(4\pi)^4m_W^2}\sum_\beta\sum_{i,j}
\sqrt{x_ix_j}
\Bigl[
J_{ij\alpha\beta}^{M}\ 
I_M'\left(x_i,x_j,x_\alpha,x_\beta\right) 
+J_{ij\alpha\beta}^{D} \ I_D'\left(x_i,x_j,x_\alpha,x_\beta\right)
\Bigr]\nonumber\\
\hspace{-0.2cm}&\approx&\hspace{-0.2cm}
-\frac{g_2^4\ e\ m_\alpha}{4(4\pi)^4m_W^2}\sum_\beta\sum_{i,j}
\sqrt{x_ix_j}
\Bigl[
J_{ij\alpha\beta}^{M}\ 
I_M'\left(x_i,x_j\right)+J_{ij\alpha\beta}^{D}\  I_D'\left(x_i,x_j\right)
\Bigr]\ ,
\label{eq:edm2}
\end{eqnarray}
where the relations among the loop functions are simply given by 
\begin{eqnarray}
I_{M,D}=\sqrt{x_ix_j}\ I_{M,D}', \  \text{and}\ \
 I'_{M,D}(x_i,x_j)\equiv I'_{M,D}(x_i,x_j,0,0)\ . 
\label{eq:edm_loop}
\end{eqnarray}

\subsection{3+1 model}
As a first  step, we consider the minimal extension of the SM with one sterile fermion state, the 
$3+1$ model,  which can be seen as the simplest effective model
mimicking extensions of the SM accounting for neutrino data.  
Since the active neutrino masses and the charged lepton
masses are much smaller than the $W$ boson mass, the loop functions are 
expanded to first order in terms of $x_{\beta=e,\mu,\tau}$ and $x_{i=1,2,3}$. 
Taking into account the fact that $\sum_{i}U_{\alpha i}^*m_iU_{\beta
i}^*=0$ holds,  one has in the 3+1 model
\begin{equation}
\sum_{i=1}^3\sqrt{x_i}J_{i4\alpha\beta}^M=0\, ,
\end{equation} 
and the EDM formula of Eq.~(\ref{eq:edm2}) reduces to 
\begin{equation}
d_\alpha\approx-\frac{g_2^4\ e\ m_\alpha}{2(4\pi)^4m_W^2}
\sum_\beta\sum_{i=1}^3
\sqrt{x_ix_4}
\left[
J_{i4\alpha\beta}^M\ x_i\frac{\partial I_M'}{\partial x_i}\left(0,x_4\right)
+J_{i4\alpha\beta}^D\ I_D'\left(0,x_4\right)
\right].\label{eq:edm2:3+1}
\end{equation}
The first term in Eq.~(\ref{eq:edm2:3+1}) is highly suppressed by the factor
$x_i=m_i^2/m_W^2\sim10^{-24}$ for $m_i\sim0.1~\mathrm{eV}$ (notice that
the sum over $i$ runs over the three light - mostly active - neutrino
masses) and thus the 
EDM is dominated by the second term. Nevertheless,  we have numerically checked that this latter term gives
a negligible contribution to the EDMs. 
For instance, assuming $\mathcal{O}(1)$ loop 
functions and the mass of the sterile fermion state, e.g., $m_4\sim
m_W$, the predicted EDM for the  tau charged lepton is 
$|d_\tau|/e\lesssim10^{-35}~\mathrm{cm}$. 
The EDMs for the  electron and the muon are even  smaller since the EDM is
proportional to the charged lepton mass (see Eq.~(\ref{eq:edm2:3+1})). 
Therefore, one can conclude that in the simple extension with one
sterile neutrino, the predicted EDMs for the charged leptons are far
below any future sensitivity. We
thus consider the next to minimal extension of the SM by two sterile
fermion states with masses $m_4$ and $m_5$.

\subsection{3+2 model}\label{sec:3+2}
The $3+2$ model is considered here as the next simplest
effective model, where the masses of the two additional sterile states
are greater that the ones of the active neutrinos. 
Taking this into
account, and the fact that the phase factors and the loop functions are both 
anti-symmetric under the exchange $i\leftrightarrow j$ (see the
discussion in Section~\ref{steps}), the EDM formula in
Eq.~(\ref{eq:edm2}) can be reduced as follows
\begin{equation}
d_\alpha\approx
-\frac{g_2^4\ e\ m_\alpha}{2(4\pi)^4m_W^2}
\sqrt{x_4x_5}\ 
\Bigl[J_{\alpha}^M\ 
I_M'\left(x_4,x_5\right)+
J_{\alpha}^D\ 
I_D'\left(x_4,x_5\right)\Bigr],
\label{eq:edm_2eff}
\end{equation}
where $J_\alpha^{M,D}$ is defined by
\begin{eqnarray}
J_\alpha^{M,D}
\hspace{-0.2cm}&\equiv&\hspace{-0.2cm}
\sum_{\beta}J_{45\alpha\beta}^{M,D}\ .
\label{ji3+2}
\end{eqnarray}
From this formula of Eq. (\ref{eq:edm_2eff}), one can see that the predicted contribution to
the EDM has the potential to be large in this model if the scale of the
two sterile neutrino masses is larger than the electroweak one
($x_{4,5}\gtrsim1$). 
The expressions for the reduced loop functions $I_M'$ and $I_D'$
(cf. Eq.~(\ref{eq:edm_loop})) are obtained for the $3+2$ model; the
expressions of the dominant parts are provided in the Appendix.

If one considers further additional sterile fermion states, all the
above discussion remains valid and we expect larger values for the
charged lepton EDMs when the extra sterile states are heavier than the
$W$ gauge boson mass. 

In the numerical analysis we will consider this  minimal  
scenario with only two sterile neutrinos (taking their masses, their
active-sterile mixing angles and  all the CP-violating phases as free
parameters). 

%%%%%%%%%%%%%%%%%%%%%%%%%%%%%%%%%%%%%%
\section{Constraints}\label{sec:constraints}

The modifications of the vertices in Eq.~(\ref{eq:lag}) due to the
presence of the rectangular $3\times (3+2)$ leptonic mixing matrix
imply deviations from unitarity of the ($3\times 3$) PMNS mixing
matrix; moreover having massive sterile neutrinos as final decay
products can possibly induce further 
deviations from the SM theoretical expectations. Consequently, scenarios
with sterile fermions are severely constrained by numerous observables and bounds, among them  EW precision measurements, charged lepton flavour
violating  observables and collider data. 

 In the following we address the most stringent constraints on the $3+2$ model. 
 We focus on sterile neutrinos with masses above the  GeV 
 since the charged lepton EDMs roughly increase with the sterile
 neutrino mass as we have discussed in the previous section.  

%%%%%%%%%%%%%%%
\subsection{Lepton flavour violating processes}

Charged lepton flavour violating  processes such as $\mu\to e\gamma$ and
$\mu\to e\overline{e}e$ give strong constraints on the parameter
space. 
The radiative process $\mu\to e\gamma$ is induced at the one-loop level, and its
 branching ratio is computed as~\cite{Abada:2015zea}
\begin{equation}
\mathrm{Br}(\mu\to e\gamma)=\frac{\sqrt{2}G_F^2m_\mu^5}{\Gamma_\mu}
\left|\sum_{i=4}^{3+N}U_{\alpha i}U^*_{\beta i}G_\gamma\left(\frac{m_i^2}{m_W^2}\right)\right|^2,
\end{equation}
where $G_F$ is the Fermi constant, $\Gamma_\mu$ is the muon total decay
width given by~\cite{Ilakovac:1994kj}
\begin{equation}
\Gamma_\mu=\frac{G_F^2m_\mu^5}{192\pi^3}\left(1-8\frac{m_e^2}{m_\mu^2}\right)
\left[
1+\frac{\alpha_\mathrm{em}}{2\pi}\left(\frac{25}{4}-\pi^2\right)
\right],
\end{equation}
and the loop function $G_\gamma(x)$ is given by~\cite{Ilakovac:1994kj}
\begin{equation}
G_\gamma(x)=\frac{x-6x^2+3x^3+2x^4-6x^3\log{x}}{4(1-x)^4}.
\label{eq:ggamma}
\end{equation}
The current experimental bound for this process is $\mathrm{Br}(\mu\to
e\gamma)\leq 5.7\times10^{-13}$~\cite{Adam:2013mnn},
and the expected future sensitivity by the upgraded MEG experiment is
$\mathrm{Br}(\mu\to e\gamma)\sim6\times10^{-14}$~\cite{Baldini:2013ke}. 

Another cLFV process, $\mu\to e\overline{e}e$,  is also induced at the one-loop
level and its branching ratio is given by~\cite{Ilakovac:1994kj}
\begin{eqnarray}
\mathrm{Br}(\mu\to e\overline{e}e)
\hspace{-0.2cm}&=&\hspace{-0.2cm}
\frac{G_F^4m_W^4m_\mu^5}{6144\pi^7\Gamma_\mu}
\left[
2\left|\frac{1}{2}F_{\mathrm{Box}}^{\mu eee}+F_Z^{\mu
  e}-2\sin^2\theta_W\left(F_Z^{\mu e}-F_{\gamma}^{\mu e}\right)\right|^2\right.\nonumber\\
\hspace{-0.2cm}&&\hspace{-0.2cm}
\left.+4\sin^4\theta_W\left|F_Z^{\mu e}-F_\gamma^{\mu e}\right|^2
+16\sin^2\theta_W\mathrm{Re}\left\{\left(F_Z^{\mu
				   e}+\frac{1}{2}F_{\mathrm{Box}}^{\mu
				   eee}\right)G_{\gamma}^{\mu e*}\right\}
\right.\nonumber\\
\hspace{-0.2cm}&&\hspace{-0.2cm}
\left.
-48\sin^4\theta_W\mathrm{Re}\left\{
\left(F_Z^{\mu e}-F_\gamma^{\mu e}\right)G_\gamma^{\mu e*}
\right\}
+32\sin^4\theta_W\left|G_\gamma^{\mu e}\right|^2
\left\{
\log\frac{m_\mu^2}{m_e^2}-\frac{11}{4}
\right\}
\right],\nonumber\\
\end{eqnarray}
where the relevant loop functions $F_\mathrm{Box}^{\mu eee}$, $F_Z^{\mu
e}$, $F_\gamma^{\mu e}$ and $G_\gamma^{\mu e}$ are given in
Ref.~\cite{Alonso:2012ji}. 
The experimental bound for this process is $\mathrm{Br}(\mu\to
e\overline{e}e)\leq1.0\times10^{-12}$~\cite{Bellgardt:1987du}. 
According to the research proposal of the Mu3e experiment, the
sensitivity will reach $\mathrm{Br}(\mu\to e\overline{e}e)\sim10^{-16}$~\cite{Blondel:2013ia}. 

While the other constraints, which will be discussed below, are mainly
related to the  mixings $|U_{\alpha i}|^2$, the cLFV processes constrain
combinations like $|U_{\alpha i}U_{\beta i}^*|$. 
In our analysis, we have computed the above mentioned observables in the
3+2 model.

%%%%%%%%%%%%%%%
\subsection{Direct collider production}
When the sterile neutrino mass is
$m_i\lesssim\mathcal{O}(100)~\mathrm{GeV}$, a strong constraint is
given by the LEP experiment. The relevant process is
$e^+e^-\to\nu_i\nu_j^*\to\nu_ie^\pm W^\mp$ where $i\leq3$ and 
$j\geq4$ and it violates lepton number conservation due to Majorana
neutrinos. 
Hence, certain regimes of the mixing angles $|U_{\alpha i}|$ are already excluded by
LEP data~\cite{Deppisch:2015qwa}. 

The other bound is given by LHC data for
$m_i\gtrsim\mathcal{O}(100~\mathrm{GeV})$ searching for a same sign
di-lepton 
channel $pp\to {W^\pm}^*\to\ell^{\pm}\nu_i\to\ell^\pm\ell^\pm jj$ where
$i\geq4$ and $j$ denotes a jet. 
With an  integrated luminosity of $20~\mathrm{fb}^{-1}$ at  
$\sqrt{s}=8~\mathrm{TeV}$, LHC data allows to constrain~\cite{Dev:2013wba,Das:2014jxa} the mixing angle $|U_{\alpha i}|$
for  sterile neutrino masses up to $500~\mathrm{GeV}$. 

As future prospects, the $\sqrt{s}=14~\mathrm{TeV}$ LHC~\cite{Dev:2013wba,Das:2015toa} and Future Lepton Colliders (like ILC)~\cite{Banerjee:2015gca,Das:2012ze} are
expected to give a stronger bound on the mixing angles. 
The collider bounds discussed here including the other constraints have
been investigated in detail in Refs.~\cite{Atre:2009rg, Deppisch:2015qwa, Alekhin:2015byh}.

\subsection{Electroweak precision data}
The active-sterile mixings  affect 
electroweak precision observables such as the $W$ boson decay width,
the $Z$ invisible decay, meson decays and the non-unitarity of the
$3\times3$ sub-matrix  ($U_{\text{PMNS}}$) of $U_{ij}$.

The constraints on the $W$ decay and $Z$ invisible decay are mostly 
relevant for $m_i<m_{W},m_Z$, respectively. 
The constraints given by  DELPHI~\cite{Abreu:1996pa} and
 L3~\cite{Adriani:1992pq} Collaborations are the strongest for the sterile
neutrino mass range $3~\mathrm{GeV}\lesssim m_i\lesssim 90~\mathrm{GeV}$. 
A future high luminosity $Z$ factory, such as FCC-ee,  will give  significant
improvements for the constraints in this mass range~\cite{Blondel:2014bra,
Abada:2014cca}.

The existence of sterile neutrinos may also violate  lepton flavour
universality of meson decays such as $\pi^+\to \ell_\alpha^+\nu_\alpha$ and
$K^+\to \ell_\alpha^+\nu_\alpha$~\cite{Abada:2013aba,
Abada:2012mc, Asaka:2014kia}.
In particular, when the sterile neutrino mass is below the threshold $m_i \leq
m_{\pi^+}$, $m_{K^+}$, the meson decay gives a strong constraint on the
mixing matrix $U_{\alpha i}$ where $m_{\pi^+}=139.6~\mathrm{MeV}$ and
$m_{K^+}=493.7~\mathrm{MeV}$. 

%%%%%%%%%%%%%%%
The non-unitarity of the $U_{\text{PMNS}}$ sub-matrix is constrained by some experiments such
as the above electroweak precision data and cLFV processes. 
The constraints for each component of the sub-matrix have been discussed in
Ref.~\cite{Antusch:2008tz, Antusch:2014woa}.

%%%%%%%%%%%%%%%
\subsection{Perturbative unitarity bound}
Any coupling of the sterile fermions to the SM particles must be
perturbative; in particular, all the couplings in Eq.~(\ref{eq:lag})
should be perturbative. 
If the additional two sterile fermion states are heavy enough to 
decay into a $W$ boson and a charged lepton, or into an active neutrino
and either a $Z$ or a Higgs boson, their decay widths should comply with
the perturbative unitarity condition~\cite{Chanowitz:1978mv,Durand:1989zs,Korner:1992an,
Bernabeu:1993up,Fajfer:1998px,Ilakovac:1999md,Abada:2014cca}. 
In this case, since the dominant decay mode of the sterile neutrinos
$\nu_{4,5}$ would be $\nu_i\to\ell_\alpha^\mp W^\pm$, 
 the decay width of $\nu_i~(i=4,5)$ has to comply with the perturbative
 unitary bound\footnote{Another common criterion of perturbativity is
 that the couplings  should be less than $\sqrt{4\pi}$. This
 criterion also gives a bound similar to  Eq.~(\ref{eq:pert}).}:
\bea 
\frac{\Gamma_{\nu_i}}{m_{i}}\, < \, \frac{1}{2}\quad \text{where} \quad \Gamma_{\nu_i}\approx
\frac{g_2^2 m_i^3}{16\pi m_W^2}\sum_{\alpha}\left|U_{\alpha
i}\right|^2\ , \, (i =4,5)\ , \
\eea
which translates into an upper bound on the sterile neutrino masses as follows, 
\begin{equation}
m_i\lesssim873~\mathrm{GeV}
\left(\sum_{\alpha}|U_{\alpha i}|^2\right)^{-1/2}.
\label{eq:pert}
\end{equation}

%%%%%%%%%%%%%
\subsection{Other constraints}
Cosmological observations, see for instance Ref.~\cite{Kusenko:2009up}, 
 put severe constraints on sterile neutrinos with a mass below the GeV
 scale. Since, and as already discussed, the contributions of the
 sterile fermion states to charged lepton EDMs are  negligible for masses below the EW
 scale, we do not apply constraints from Big Bang Nucleosysthesis or
 Cosmic Microwave Background, which would be
 relevant in the very low mass regime. 
  For heavier mass regimes,  the perturbative unitarity condition of
 Eq.~(\ref{eq:pert})  turns out to be also very constraining.

The additional sterile fermions might contribute to neutrinoless double
beta decay, and the corresponding effective mass  $m_{ee}$ is corrected
according to~\cite{Blennow:2010th,Abada:2014nwa}
\begin{equation} 
\label{eq:0vudbdecay}
 m_{ee}\,\simeq \,\sum_{i=1}^{5} U_{ei}^2 \,p^2
\frac{m_i}{p^2-m_i^2}\ , \end{equation}
where $p^2 \simeq - (125 \mbox{ MeV})^2$ is the virtual momentum of the neutrino. 
Several experiments (among them GERDA~\cite{Agostini:2013mzu}, 
EXO-200~\cite{Auger:2012ar,Albert:2014awa},
KamLAND-ZEN~\cite{Gando:2012zm}) have put constraints on the effective
mass, which translate into bounds on combinations of $U_{ei}^2 m_i$,
$i=4, 5$. In our numerical analysis, we have checked that our solutions
always comply with the conservative experimental bound $|m_{ee}|
\lesssim 0.01$ eV (since the cancellation between each $i$-th contribution
can occur due to the existence of the CP-violating phases).

%%%%%%%%%%%%%%%%%%%%%%%%%%%%
\section{Numerical Results}\label{Sec:Numerical Results}
We now proceed with the numerical evaluation of the  EDMs of charged
leptons in the $3+2$ model. 
As described in Section \ref{sec:model}, this simple model allows to
illustrate the potential effects of the addition of the number $N>2$ of
sterile fermion states. 
For completeness, we also address the muon magnetic dipole moment, the
first non-vanishing contribution arising at the one-loop level. 

In the numerical analysis we have conducted, we consider the minimal
scenario with only two sterile neutrinos, taking their masses, their
active-sterile mixing angles and  all the CP-violating phases as free
parameters. 
\subsection{Electric dipole moments of charged leptons}
In order to investigate the parameter space of the $3+2$ model for the
charged lepton EDMs, we vary the  parameters in the following  ranges: 
\begin{equation}
1~\mathrm{GeV}\leq m_i\leq 10^6~\mathrm{GeV},\quad
\sin\theta_{j i}\leq 0.1,\  \text{for } j =1,2,3 \  \text{and } i=4,5,\quad
\sin\theta_{45}\leq1,
\end{equation}
and all the Dirac and Majorana CP-violating phases are taken in the range of $[0,2\pi]$. 
The mixing between the two sterile states, $\theta_{45}$, is not constrained. 

\begin{figure}[t]
\begin{center}
\hspace*{-0.5cm}
\includegraphics[scale=0.64]{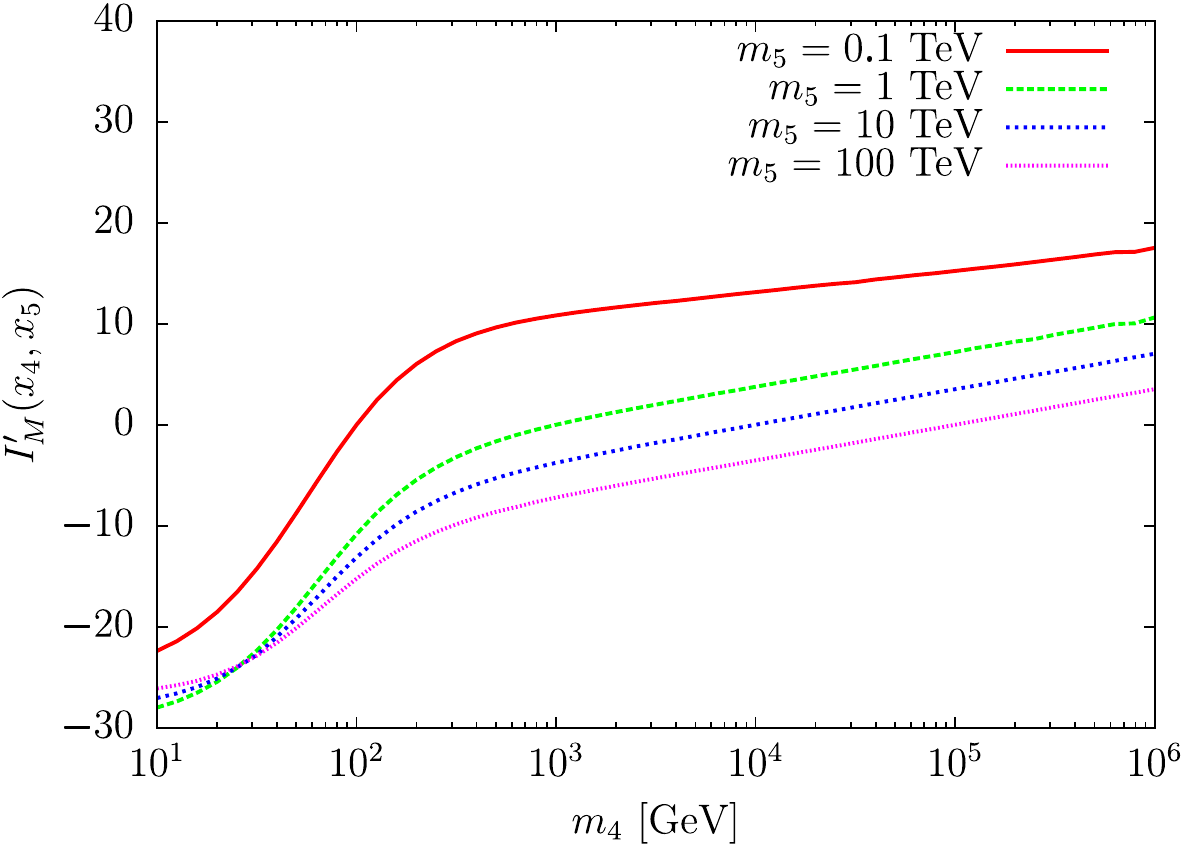}
\hspace*{0.3cm}
\includegraphics[scale=0.64]{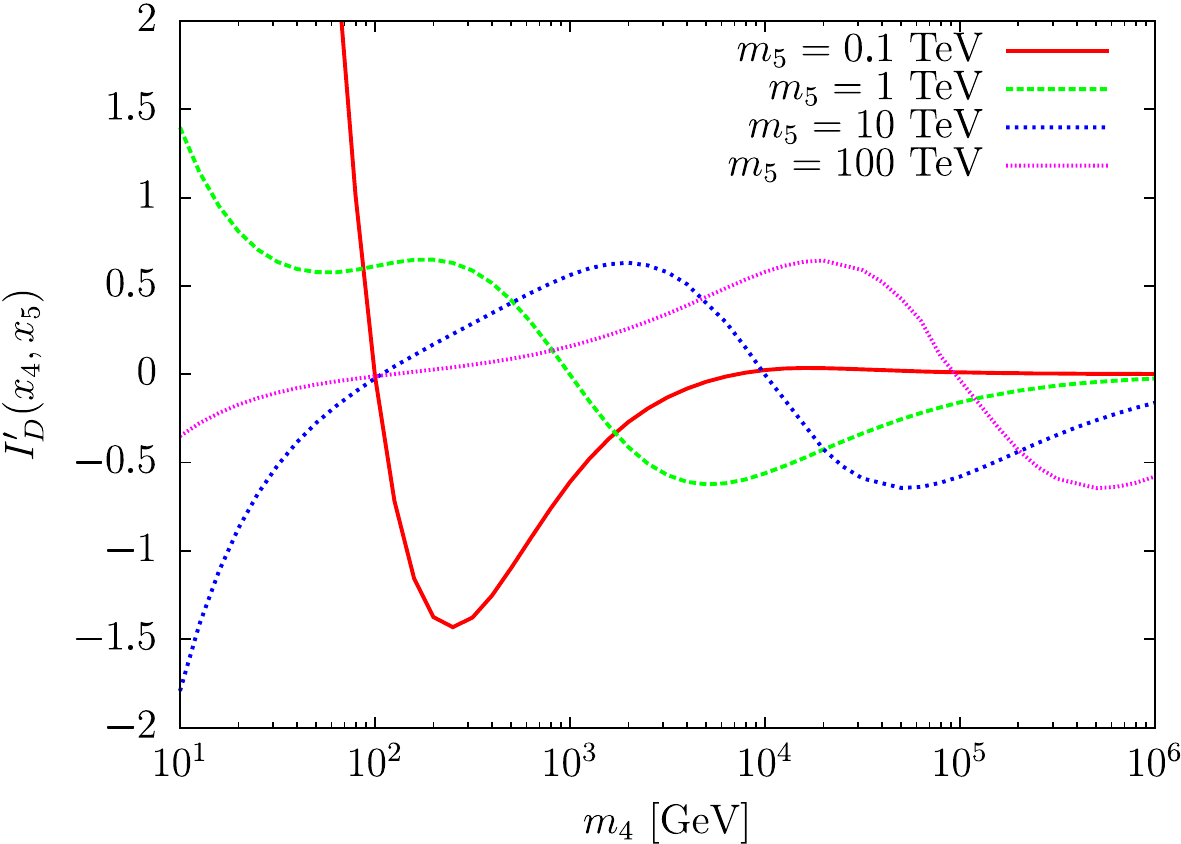}
\caption{Loop functions $I_M'$ and $I_D'$ as a function of $m_4$ for
 several fixed values of $m_5$.}
\label{fig:loop_f}
\end{center}
\end{figure}

The reduced loop functions $I_M'$ and $I_D'$, defined in
Eq.~(\ref{eq:edm_2eff}), and whose analytical expressions of the dominant
parts are given in the Appendix, have been 
numerically evaluated. To illustrate their relative contributions (their
coefficients, namely the phase factors $J_\alpha^{M,D}$, saturating at
1), we display on Fig.~\ref{fig:loop_f} the loop integrals $I_M'$ and
$I_D'$ as a function of $m_4$ for several fixed  values of $m_5$. 
One can see that the loop function $I_D'$ is
$\mathcal{O}(1)$ in 
most of the considered mass range, while the loop function
$I_M'$ increases logarithmically with $x_4$. 
Due to this, the Majorana contribution $J_\alpha^MI_M'$ in
Eq.~(\ref{eq:edm_2eff}) gives the  dominant contribution
to the EDMs in most of the considered mass range, independently of the
active-sterile mixings.

\begin{figure}[t]
\begin{center}
\includegraphics[scale=0.6]{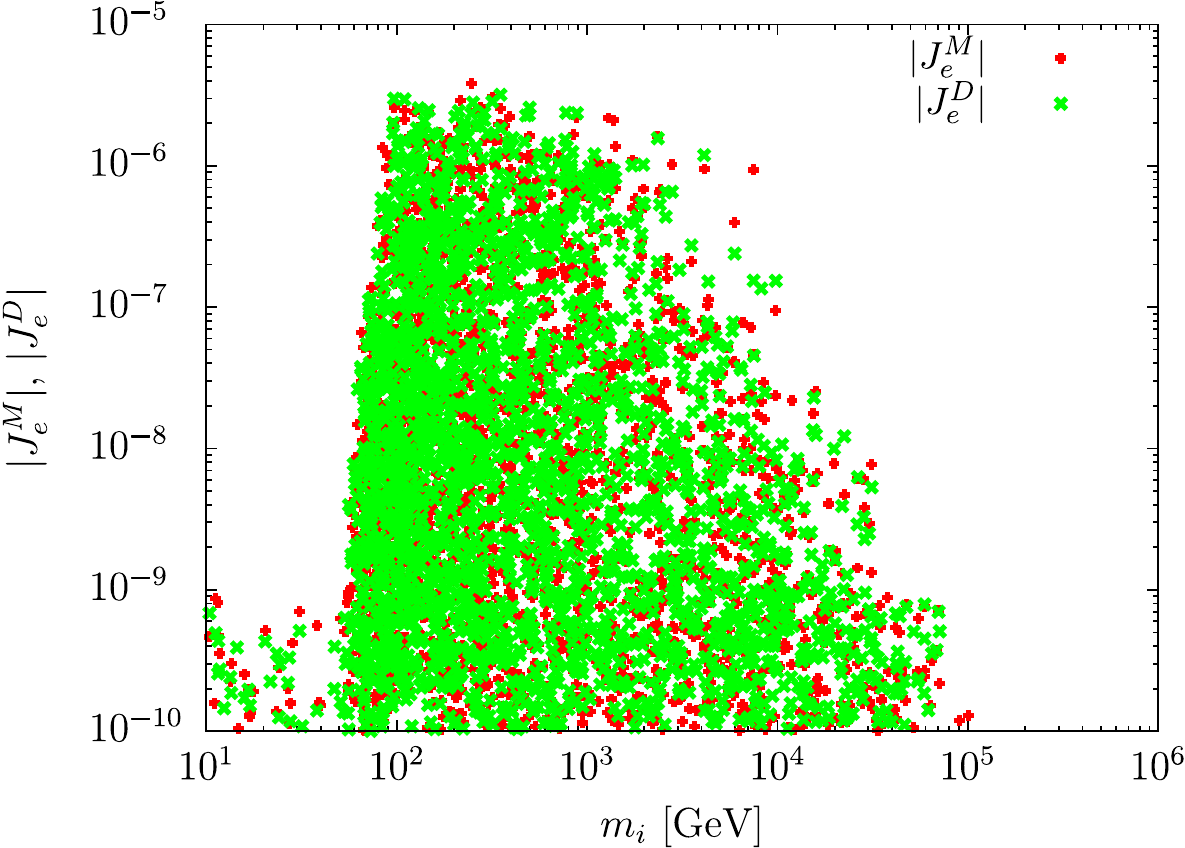}
\hspace{0.1cm}
\includegraphics[scale=0.6]{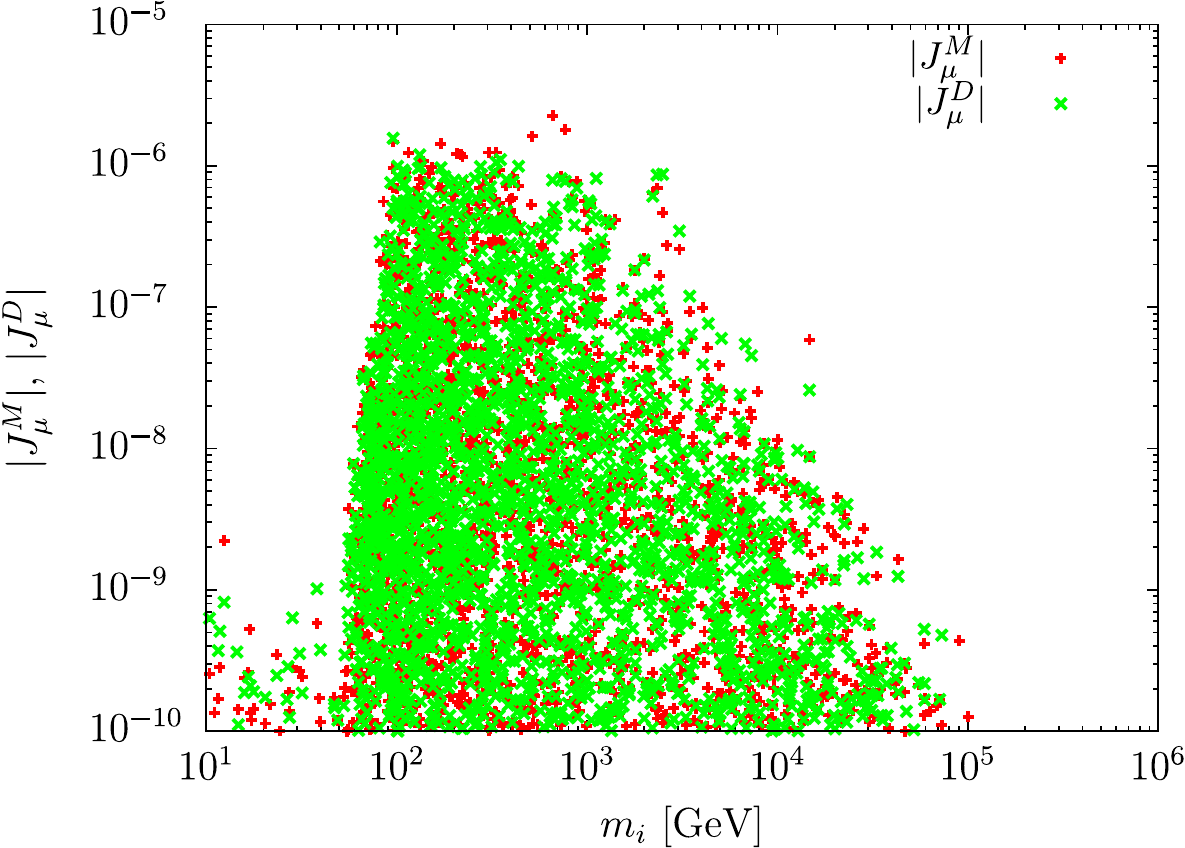}\\
\vspace{0.3cm}
\includegraphics[scale=0.6]{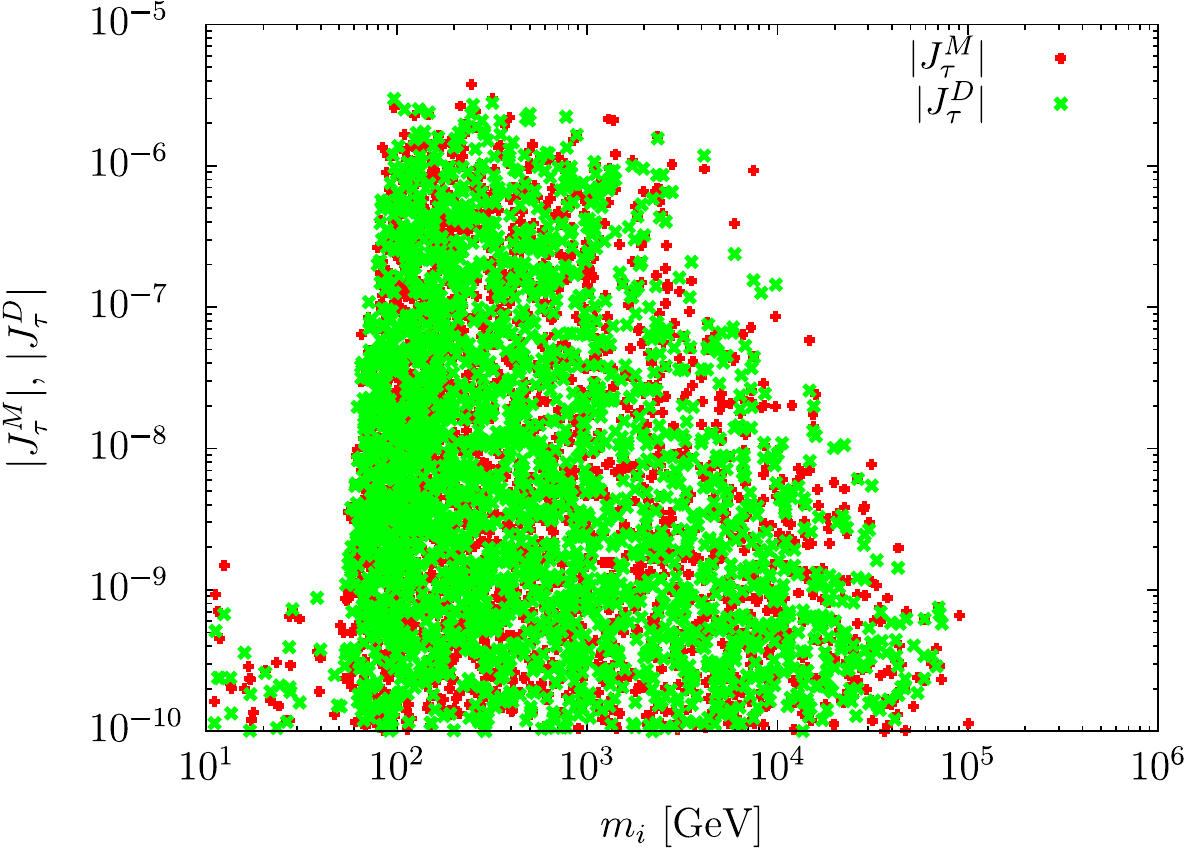}
\caption{Phase factors $|J_\alpha^M|$ and $|J_\alpha^D|$ as a function of $m_i$,  where
 {\small $\displaystyle J_\alpha^{M,D}=\sum_{\beta}J_{45\alpha\beta}^{M,D}$} for the three charged leptons.} 
\label{fig:J_factor}
\end{center}
\end{figure}

On Fig.~\ref{fig:J_factor} we display  the allowed parameter space for
the phase factors $|J_\alpha^M|$ and 
$|J_\alpha^D|$ defined in Eqs.~(\ref{eq:edm:IJ}), (\ref{ji3+2}), after
having applied all the constraints discussed in Section
\ref{sec:constraints}. 
As one can see, in each of the 3 panels corresponding to the three
charged leptons, the upper right region is excluded by the perturbative
unitarity constraint. 
The region corresponding to $m_i\lesssim70~\mathrm{GeV}$ is strongly
constrained by LEP. Constraints from electroweak precision data are also
important and almost independent of the sterile neutrino mass when
$m_i\gtrsim1~\mathrm{GeV}$. Finally, the region which would lead to  $|J_\alpha^M|,\
|J_\alpha^D|\gtrsim10^{-6}$ is constrained by the bounds on cLFV
processes. 
As one can see from this figure, there is no substantial difference
between $|J_\alpha^M|$ and $|J_\alpha^D|$. 
Moreover, the allowed region for all the charged
leptons is almost the same. 
Therefore, using Eq.~(\ref{eq:edm_2eff}), an approximate  relation among
the charged lepton EDMs is found as 
\begin{equation}
\frac{|d_e|}{m_e}\sim\frac{|d_\mu|}{m_\mu}\sim\frac{|d_\tau|}{m_\tau}\ . 
\end{equation}

\begin{figure}[t]
\begin{center}
\includegraphics[scale=0.6]{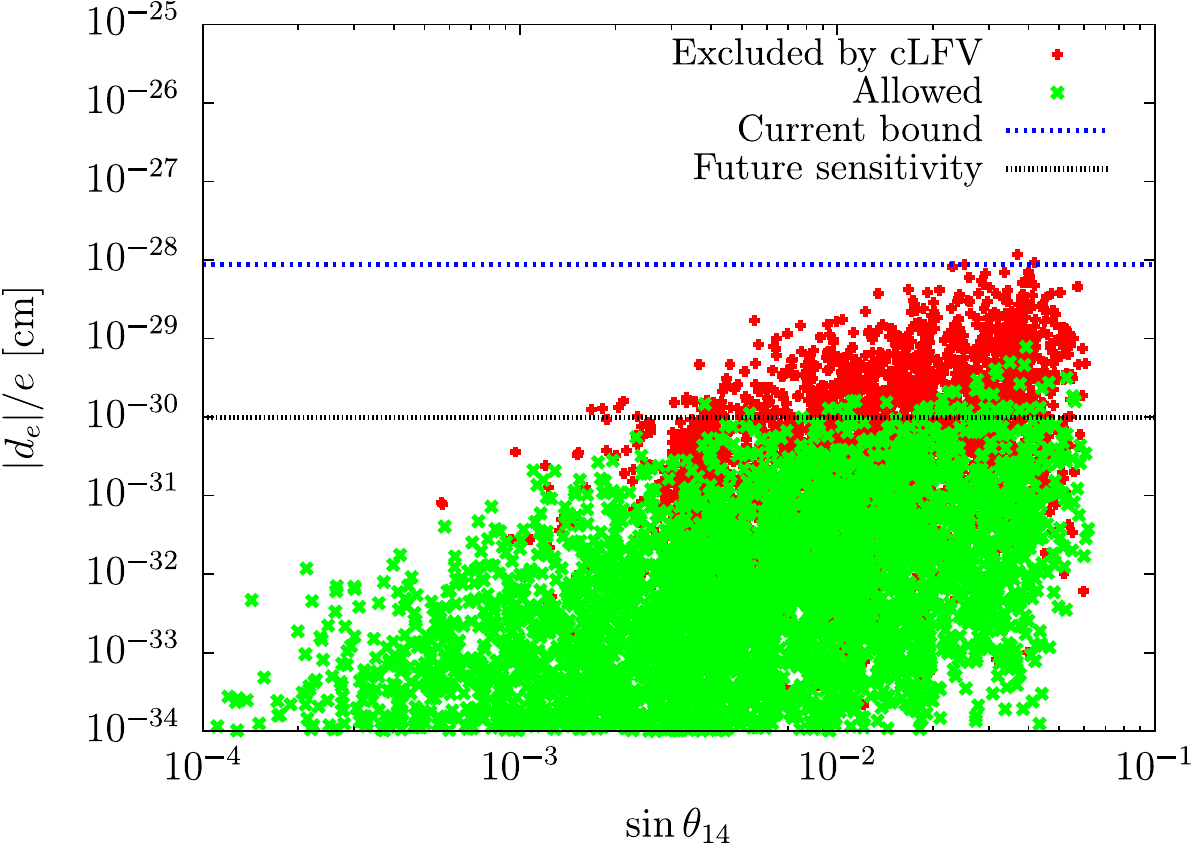}
\hspace{0.1cm}
\includegraphics[scale=0.6]{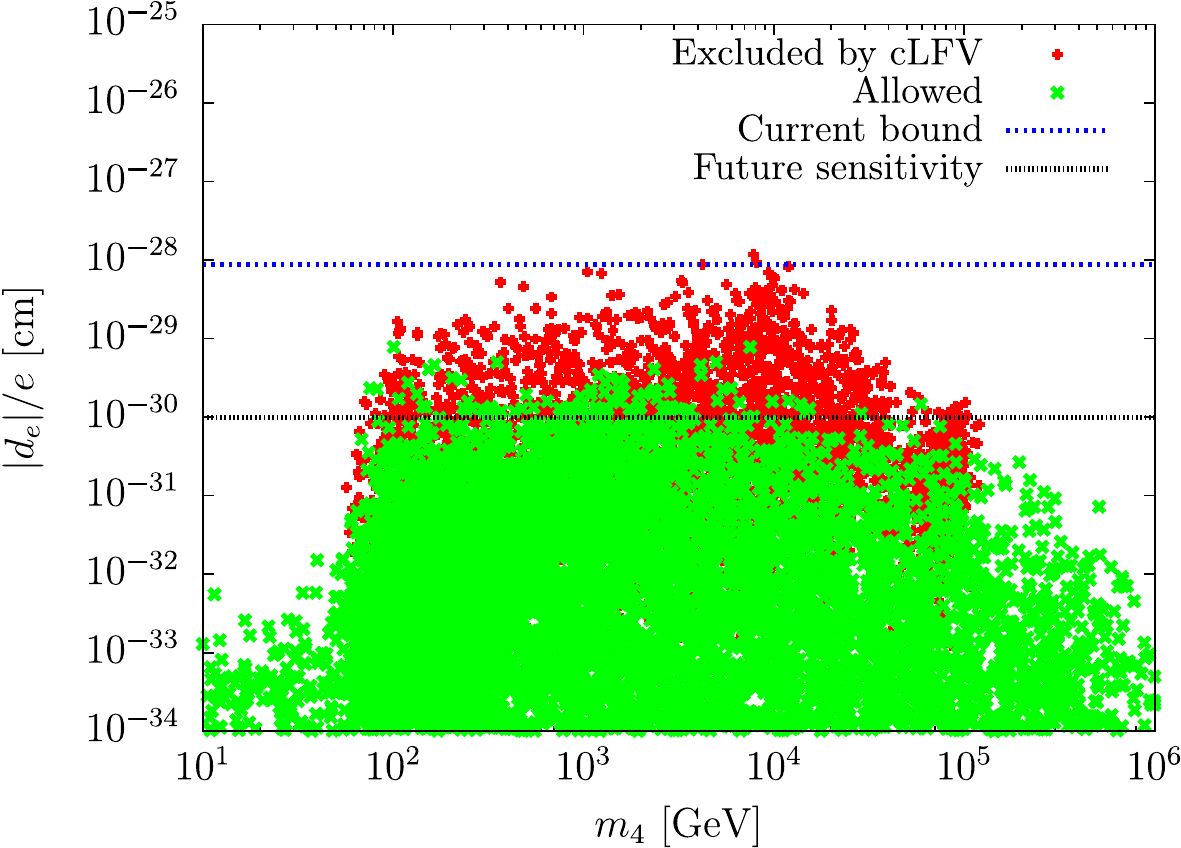}\\
\vspace{0.2cm}
\includegraphics[scale=0.6]{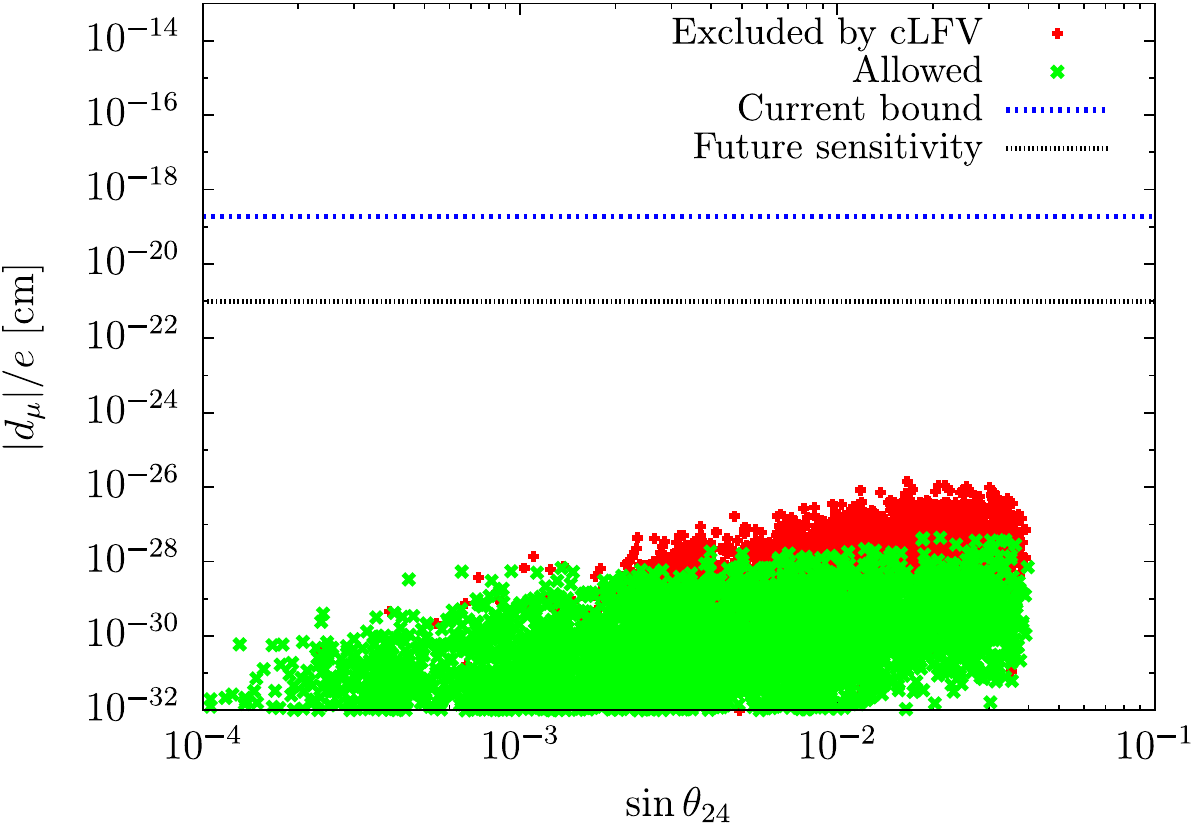}
\hspace{0.1cm}
\includegraphics[scale=0.6]{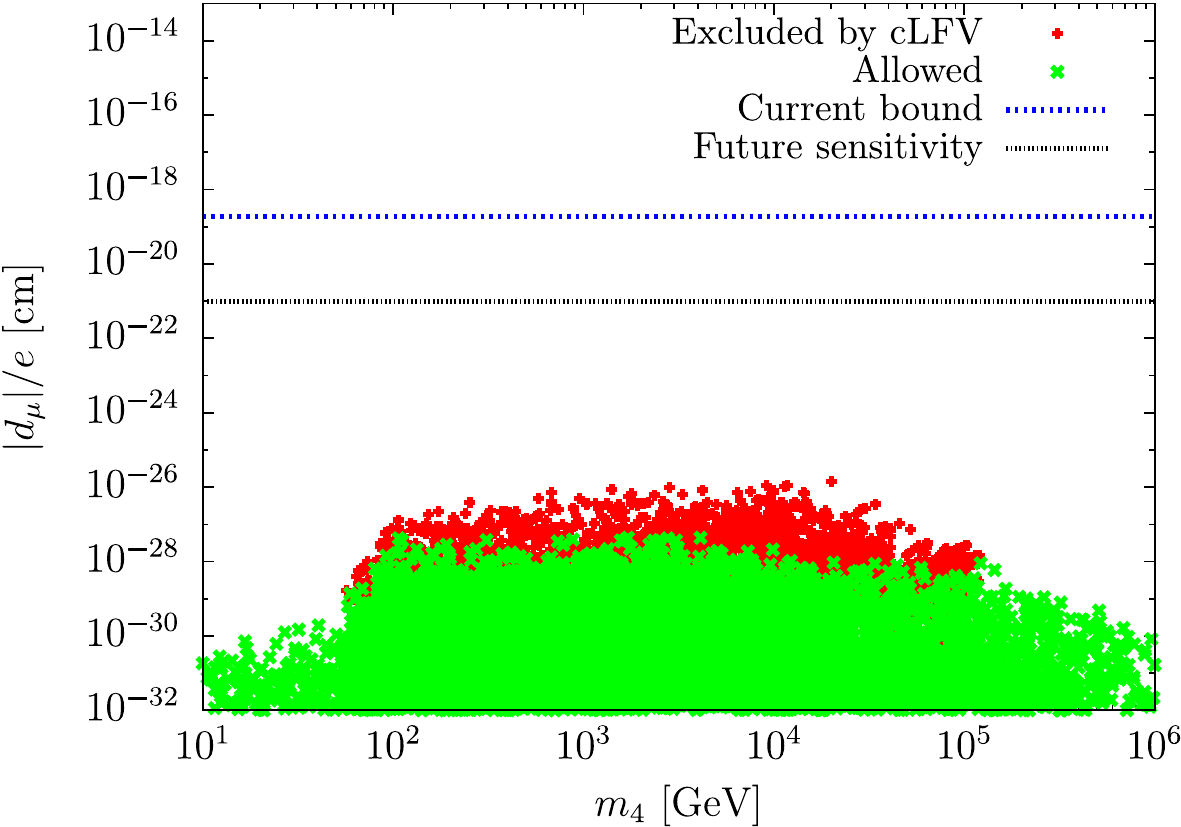}\\
\vspace{0.2cm}
\includegraphics[scale=0.6]{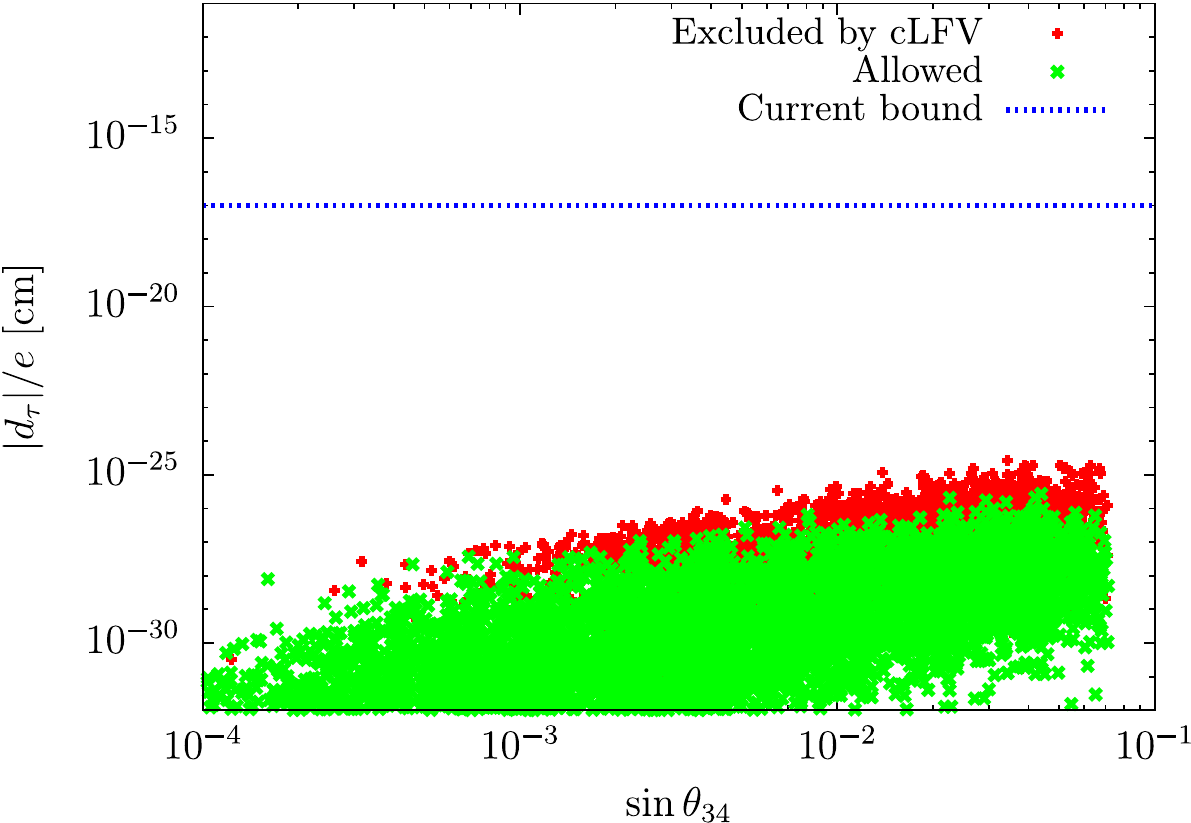}
\hspace{0.1cm}
\includegraphics[scale=0.6]{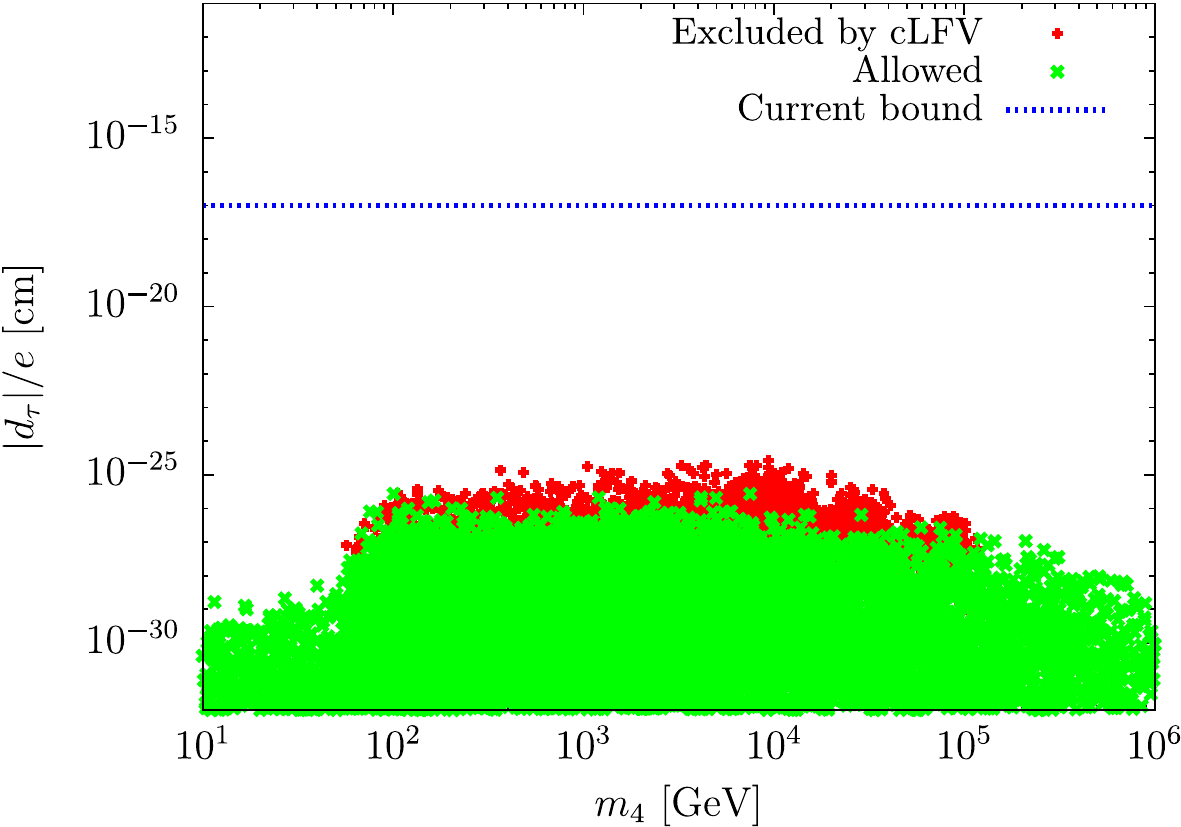}
\caption{Charged lepton EDMs in the $3+2$ effective model as a
 function of $\theta_{i4}$, $i=1, 2, 3$ (left) and $m_4$ (right). The current upper
 bounds and future prospects are also shown as blue and black lines respectively.} 
\label{fig:num_edm}
\end{center}
\end{figure}

For the theoretically and experimentally viable regions of the parameter
space, the charged lepton EDMs are shown in Fig.~\ref{fig:num_edm} as a
function of $\theta_{i4}\ , i=1,2,3$ (left panels) and of $m_4$ (right
panels), respectively. 
The green points comply with all the imposed constraints while the red points are
excluded by the bounds on  cLFV processes. 
The current experimental bounds and
future sensitivities are also shown as blue and black lines. 
Analogous results would be obtained when displaying these observables as
a function of $\theta_{i5}$ and  $m_5$, as the two sterile states play
similar r\^oles in the several diagrams. 
As one can see from Fig.~\ref{fig:num_edm}, the maximum values for the
EDMs are obtained in the ranges of
$\sin\theta_{i4}\gtrsim10^{-2}~(i=1,2,3)$ and 
$100~\mathrm{GeV}\lesssim m_4\lesssim100~\mathrm{TeV}$. 
The range  $\sin\theta_{i4}\gtrsim0.1$ is excluded  by the constraints
discussed in the previous section, 
in particular by  electroweak precision data. 
The electron EDM is always below the current experimental upper bound
$|d_e|/e<8.7\times10^{-29}~\mathrm{cm}$ (so no additional bound on the
parameter space arises from this CP-violating observable), but the
corresponding contributions can be within the optimistic 
future sensitivity, $|d_e|/e=10^{-30}~\mathrm{cm}$. 
On the other hand for the muon and tau, the predicted EDMs are much
smaller than any  future sensitivity. 

%%%%%%%%%%%%
One can also try to  investigate which combinations of the $10$ physical
CP-violating phases are relevant for the EDMs;  however, this task is
very involved and we have not identified significant correlations between
a given phase (or combination of phases) in our parametrisation and the
contributions to the EDMs.

\begin{figure}[t]
\begin{center}
\includegraphics[scale=0.6]{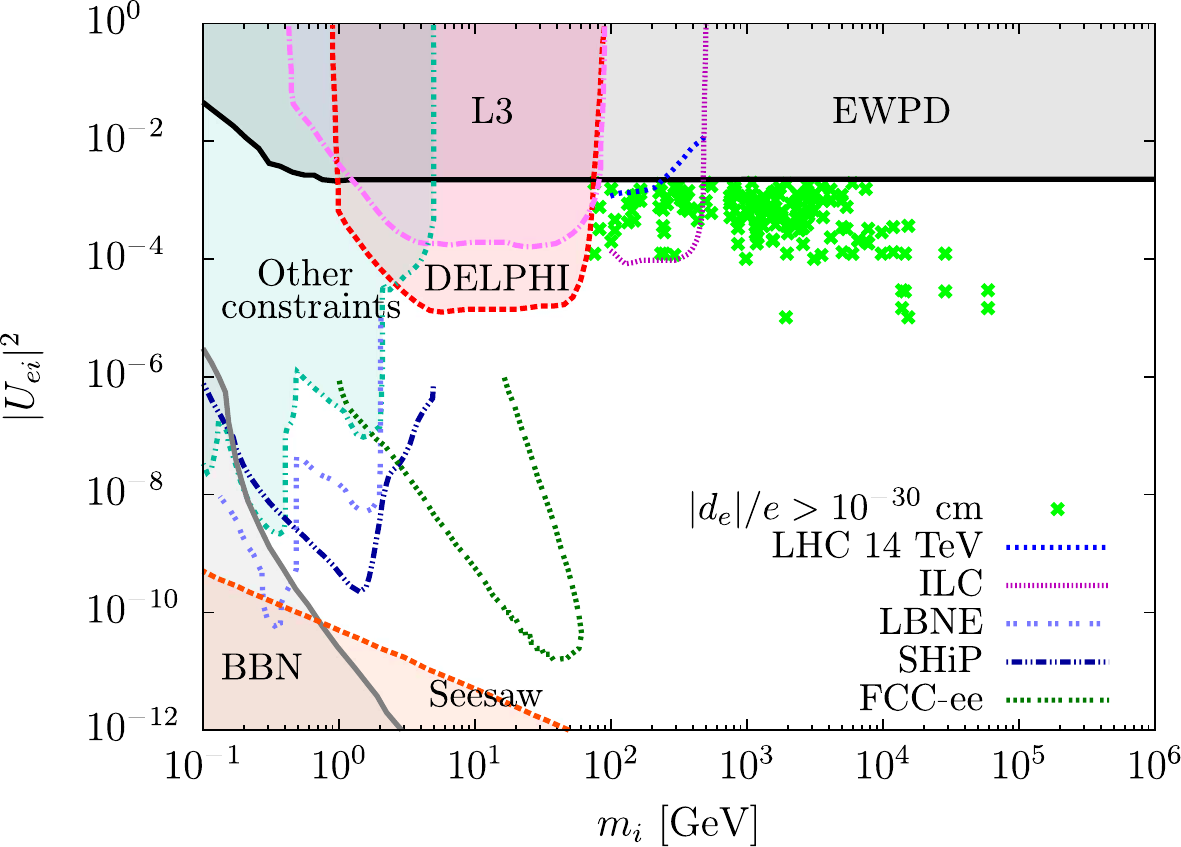}
\includegraphics[scale=0.6]{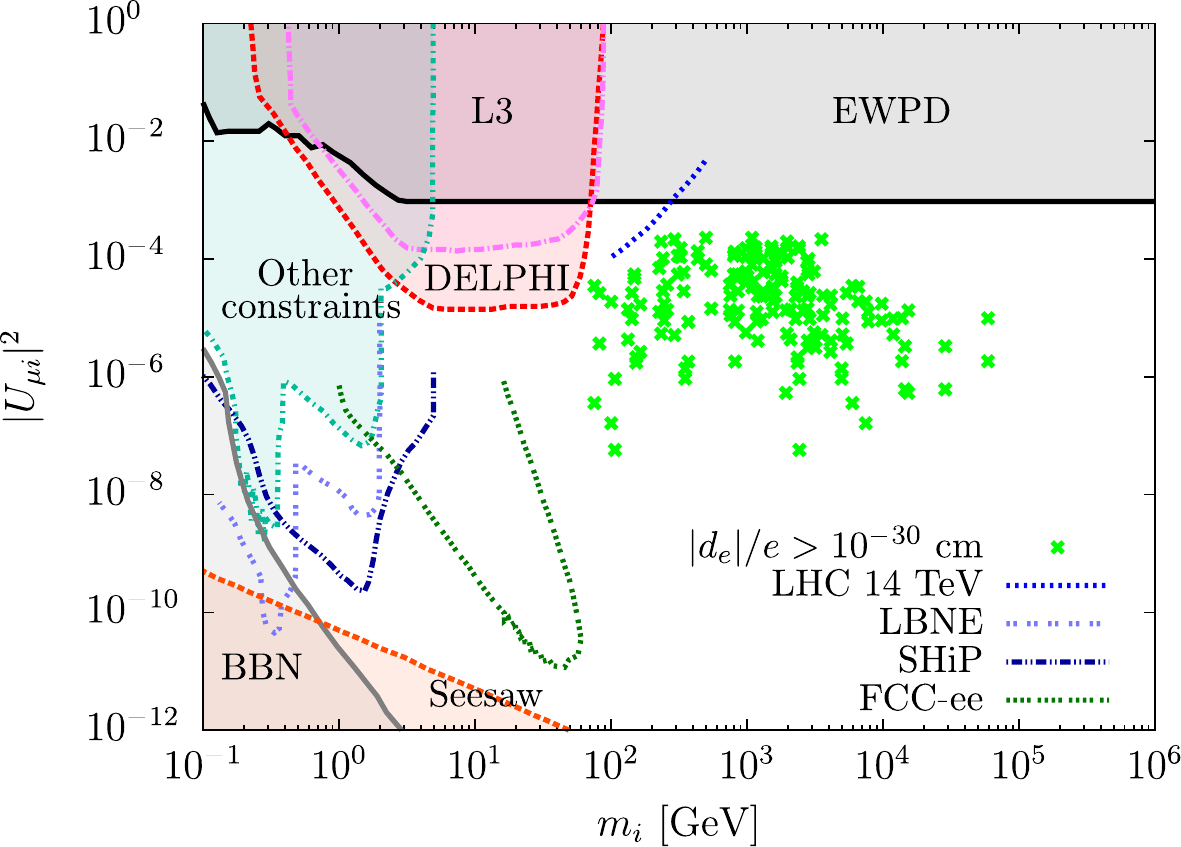}\\
\vspace{0.3cm}
\includegraphics[scale=0.6]{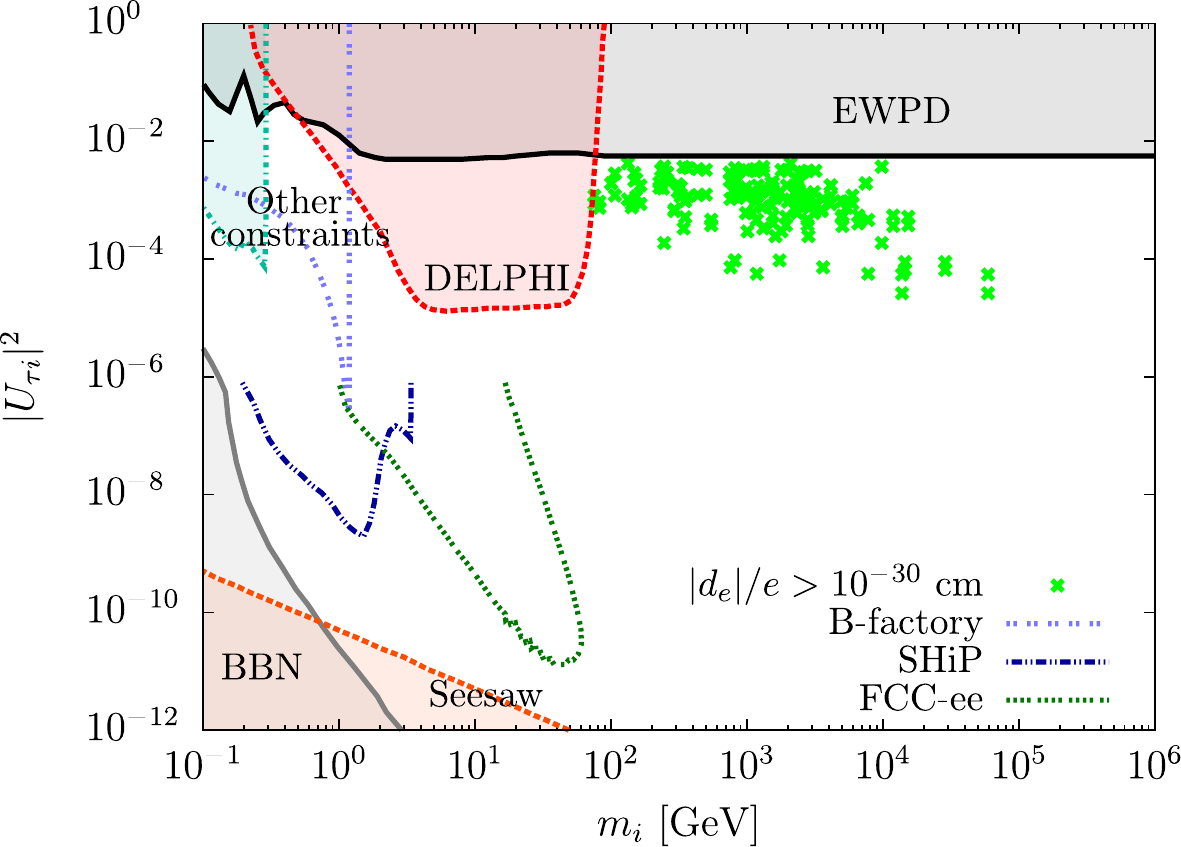}
\caption{Parameter spaces $(|U_{\alpha i}|^2, m_i)$, for $\alpha=e, \mu,
 \tau$ and $i=4$ or $5$. The coloured surfaces are excluded due to the
 violation  of at least one experimental or observational -mostly from
 BBN- bound. Lines (full, dashed and dotted)  delimit the expected sensitivity of
  several facilities: DUNE, SHiP, FCC-ee and
  LHC. Green points denote  predictions for the  
  electron EDM  within the future sensitivity reach, 
 $|d_e|/e\geq10^{-30}~\mathrm{cm}$. }
\label{fig:mi-pmns}
\end{center}
\end{figure}

To conclude the discussion of the charged lepton EDMs, we present in
Fig.~\ref{fig:mi-pmns} the predictions for the electron EDM in the
$(|U_{\alpha i}|^2, m_i)$ ($\alpha=e,\mu,\tau$, $i=4,5$) 
parameter space of the $3+ 2$ model. 
Coloured surfaces reflect the
violation of at least one phenomenological or cosmological  bound as presented  in 
Section~\ref{sec:constraints} (see also the discussion
of~\cite{Alekhin:2015byh,Deppisch:2015qwa,deGouvea:2015euy}). 
The different lines correspond to the reach of future
facilities: the projected exclusion limit from the LHC (14~TeV run data~\cite{Deppisch:2015qwa}), the
expected sensitivity of FCC-ee regarding the production of heavy 
sterile neutrinos~\cite{Blondel:2014bra}, DUNE~\cite{Adams:2013qkq} and SHiP (a 
fixed-target experiment using high-intensity proton beams at the CERN
SPS~\cite{Bonivento:2013jag,Anelli:2015pba}). 
The displayed green points correspond to having the electron EDM larger 
than the future sensitivity, i.e. $|d_e|/e>10^{-30}$.

As one can see on the first panel of Fig.~\ref{fig:mi-pmns}, in view of
the associated large regime for  the active-sterile mixing $|U_{ei}|^2$,
some points can be tested by 
future collider experiments such as the LHC with $\sqrt{s}=14$ TeV and  the future ILC. 
Despite their impressive sensitivities, future experiments such as
LBNE, SHiP and FCC-ee will not be able to probe the regions in parameter
space responsible for sizable EDM contributions, since these facilities aim 
at  sterile mass regimes below the EW scale (recall that in order to
have the electron EDM within experimental sensitivity, the sterile
masses should be  $100~\mathrm{GeV}\lesssim m_i\lesssim 
100~\mathrm{TeV}$).

%%%%%%%%%
\subsection{Muon anomalous magnetic moment}
As already mentioned, sterile fermion states can have an impact on CP- and
flavour-conserving observables. Here, and for completeness,  we briefly
address the impact of the 2 extra sterile states on the muon anomalous
magnetic moment. 
The current experimental value of the muon anomalous magnetic moment has
been measured by Muon $g-2$ Collaboration~\cite{Bennett:2006fi}, and
the discrepancy of the muon anomalous magnetic moment between the
experimental value and the SM prediction is given
by~\cite{Agashe:2014kda}
\begin{equation}
\Delta{a}_\mu\equiv
a_\mu^\mathrm{exp}-a_\mu^\mathrm{SM}=2.88\times10^{-9}.
\end{equation}
In our scenario, the muon anomalous magnetic
moment induced by the $W$ boson and neutrino loop can be computed at
one-loop level as~\cite{Abada:2014nwa} 
\begin{equation}
a_\mu=\frac{\sqrt{2}G_Fm_\mu^2}{\left(4\pi\right)^2}
\sum_{i=1}^{3+N}\left|U_{\mu
		 i}\right|^2F\left(\frac{m_i^2}{m_W^2}\right),
\label{eq:g-2}
\end{equation}
where the loop function $F(x)$ is defined by 
\begin{equation}
F\left(x\right)=
\frac{10-43x+78x^2-49x^3+4x^4+18x^3\log{x}}{3\left(1-x\right)^4}.
\end{equation}
Subtracting the (one-loop with mostly active neutrino contributing in
the loop) SM contribution in Eq.~(\ref{eq:g-2}), one obtains 
\begin{equation}
\Delta{a}_\mu\approx
-\frac{4\sqrt{2}G_Fm_\mu^2}{(4\pi)^2}\sum_{i=4}^{3+N}|U_{\mu
i}|^2G_\gamma\left(\frac{m_i^2}{m_W^2}\right),
\label{eq:mag}
\end{equation}
where the active neutrino masses $m_i~(i=1,2,3)$ are neglected and
$G_\gamma(z)$ is defined by Eq.~(\ref{eq:ggamma}). 
As  has been shown in~\cite{Abada:2014nwa}, 
the new contribution to the muon anomalous magnetic moment 
can hardly fill the unexplained discrepancy with experiment.
Taking into account all the experimental constraints discussed in
Sections~\ref{sec:constraints} and \ref{Sec:Numerical Results}, 
the predicted value of the muon anomalous magnetic moment is roughly
$\Delta{a}_\mu\sim-10^{-12}$ for $|U_{\mu i}|^2\sim10^{-3}$, and thus 
 additional contributions  to the anomalous magnetic moment are still required to
explain the discrepancy between theory and experimental measurements.

%%%%%%%%%%%%%%%%%%%%%%%%%%%%
\section{Summary}
We have discussed the contributions of sterile neutrinos to  charged
lepton EDMs in the SM minimally extended via $N$ sterile fermion
states. We have considered all the diagrams which can potentially 
contribute to charged lepton EDMs. 
The comparison of the different contributions has shown that
significant contributions to the charged lepton EDMs can only be
obtained if the (sterile) neutrinos are of Majorana nature. 

In the case of the $3+1~(N=1)$ model, we have found that the predicted
EDMs are too small to be detected in any foreseen future experiments.
We have moreover  verified that 
at least two sterile neutrinos are required to obtain an electron
EDM within future sensitivity reach. In this most minimal scenario ($N=2$),
the masses of the two sterile states should be in the range 
$100~\mathrm{GeV}-100~\mathrm{TeV}$ to have $|d_e|/e\geq10^{-30}~\mathrm{cm}$. 
For the muon and the tau, the predicted EDMs remain several orders of
magnitude below the future sensitivities.

In our analysis we imposed all available experimental and observational
constraints on sterile neutrinos, and we also discussed the prospects
of probing this scenario at low and high energy experiments. 
In particular, regions in parameter space which predict a large electron
EDM could  be also explored by collider experiments such as a Future
Linear Collider (ILC) and marginally with the 
$\sqrt{s}=14~\mathrm{TeV}$ LHC.

%%%%%%%%%%%%%%%%%%%%%%%%%%%%%%%%%%%

\section*{Acknowledgments}
The authors would like to thank Michele Lucente and Olcyr Sumensari for
fruitful discussions. A. A. is grateful to Amon Ilakovac and to Avelino Vicente for useful discussions. 
We acknowledge support from the European ITN project
(FP7-PEOPLE-2011-ITN, PITNGA-2011-289442-INVISIBLES).
T. T. acknowledges support from P2IO Excellence Laboratory (LABEX).
This work was done in the framework of a ``D\'efi InPhyNiTi'' 
project (N2P2M-SF)
%%%%%%%%%%%%%%%%%%%%%%%%%%%%%%%%%%%

%\section*{Appendix}
\section*{Loop Calculations}
In what follows we provide  the  leading terms of the loop functions in the limit where $x_i,x_j\gg1$.
When the sterile neutrino masses are much heavier than the $W$ boson mass
($x_i,x_j\gg1$), the loop functions satisfy the relation $I_M'(x_i,x_j)\gg I_D'(x_i,x_j)$
due to the additional factor $\sqrt{x_ix_j}$ for $I_M'(x_i,x_j)$ where
$I_{M,D}'(x_i,x_j)$ is defined in Eq.~(\ref{eq:edm2}). 
Moreover, the loop function $I_M'(x_i,x_j)$ contains terms increasing 
with $\log\left(x_i/x_j\right)$, which turn out to be the dominant ones. 

In general the loop function $I_M'(x_i,x_j)$ can be expressed by
anti-symmetrizing in terms of $x_i$ and of $x_j$ as
\begin{equation}
I_M'(x_i,x_j)=
\frac{1}{2}\Bigl(I_{M0}'(x_i,x_j)-I_{M0}'(x_j,x_i)\Bigr),
\end{equation}
where $I_{M0}'(x_i,x_j)$ is given by
\begin{equation}
I_{M0}'(x_i,x_j)=
{I_{M01}'}^{\hspace{-0.5cm}(a)}\hspace{0.1cm}(x_i,x_j)+
{I_{M02}'}^{\hspace{-0.5cm}(a)}\hspace{0.1cm}(x_i,x_j)+
{I_{M01}'}^{\hspace{-0.5cm}(b)}\hspace{0.1cm}(x_i,x_j)+
{I_{M02}'}^{\hspace{-0.5cm}(b)}\hspace{0.1cm}(x_i,x_j),
\label{eq:appendix}
\end{equation}
and the four functions in right-hand side in Eq.~(\ref{eq:appendix})
will be given below. The superscripts $(a),(b)$ denote contributions coming  from
the diagrams in the first and second lines in Fig.~\ref{fig:fig1}
respectively; the subscript $1,2$  translates that  two different functions exist
for each diagram. 
%%%%%%%%%%%%%%%%%%%%%
The loop function ${I_{M0n}'}^{\hspace{-0.5cm}(a,b)}$ ($n=1,2$) for the
diagrams in the first and second lines in Fig.~\ref{fig:fig1} are given by 
\begin{eqnarray}
{I_{M0n}'}^{\hspace{-0.5cm}(a)}\hspace{0.1cm}(x_i,x_j)
\hspace{-0.2cm}&=&\hspace{-0.2cm}
\int_0^1\prod_{A=1}^{4}ds_A\delta\left(\sum_{A=1}^4s_A-1\right)
\int_0^1\prod_{B=1}^{3+\delta_{2n}}dt_B\delta\left(\sum_{B=1}^{3+\delta_{2n}}t_B-1\right)
F_n^{(a)}(x_i,x_j),\quad\\
{I_{M0n}'}^{\hspace{-0.5cm}(b)}\hspace{0.1cm}(x_i,x_j)
\hspace{-0.2cm}&=&\hspace{-0.2cm}
\int_0^1\prod_{A=1}^{5}ds_A\delta\left(\sum_{A=1}^5s_A-1\right)
\int_0^1\prod_{B=1}^{3+\delta_{2n}}dt_B\delta\left(\sum_{B=1}^{3+\delta_{2n}}t_B-1\right)
F_n^{(b)}(x_i,x_j),
\end{eqnarray}
where $F_{n}^{(a,b)}(x_i,x_j)$ is given by
\begin{eqnarray}
F_n^{(a)}(x_i,x_j)
\hspace{-0.2cm}&=&\hspace{-0.2cm}
\frac{N_{n}^{(a1)}}{D^{(a1)}}
+\frac{N_{n}^{(a2)}}{D^{(a2)}}
+\frac{N_{n}^{(a3)}}{D^{(a3)}},\\
F_n^{(b)}(x_i,x_j)
\hspace{-0.2cm}&=&\hspace{-0.2cm}
\frac{N_{n}^{(b1)}}{D^{(b1)}}
+\frac{N_{n}^{(b2)}}{D^{(b2)}}.
\end{eqnarray}
Each of the terms corresponds to the contribution coming from the diagrams (a1), (a2),
(a3), (b1) and (b2). The denominators are given by
\begin{eqnarray}
D^{(a1)}=D^{(a2)}
\hspace{-0.2cm}&=&\hspace{-0.2cm}
s_2(s_2-1)(t_1x_j+t_2)-(1-t_1-t_2)(s_1x_i+s_3+s_4),\\
D^{(a3)}
\hspace{-0.2cm}&=&\hspace{-0.2cm}
(s_2+s_3)(s_2+s_3-1)(t_1x_i+t_2) - (1-t_1-t_2)(x_1x_j+x_4),\\
D^{(b1)}
\hspace{-0.2cm}&=&\hspace{-0.2cm}
-(1-t_1)(s_1+s_4+s_5+s_2x_i+s_3x_j),\\
D^{(b2)}
\hspace{-0.2cm}&=&\hspace{-0.2cm}
t_1(s_4+s_5)(s_4+s_5-1)-(1-t_1)(s_1+s_2x_i+s_3x_j),
\end{eqnarray}
and the numerators for the diagram (a1), (a2), (a3) are given by
\begin{eqnarray}
N_{1}^{(a1)}+N_{1}^{(a2)}
\hspace{-0.2cm}&=&\hspace{-0.2cm}
3 (s_1-s_2) t_2+s_1 \left(5 t_1+1\right) t_3+5 s_2
\left(1-t_1\right) t_2\nonumber\\
&&\hspace{-0.2cm}
-s_1-11 t_2-t_3+9+\frac{s_1 \left(-5 t_2-5
	       t_3+13\right) t_3}{s_2-1}\nonumber\\
&&\hspace{-0.2cm}
+\frac{x_i}{2}\frac{t_3 (s_3+s_4)}{s_2-1}
+\frac{x_j}{2} \left(\frac{5 s_1 t_3}{s_2-1}
-6\left(s_1t_3+s_2 t_2-s_1\right)+t_3-2\right),
\end{eqnarray}
%%%
\begin{eqnarray}
N_{1}^{(a3)}
\hspace{-0.2cm}&=&\hspace{-0.2cm}
-\left(5 s_1+1\right) \left(1-t_1\right)-\left(s_1+1\right)
t_2+\left(s_2+s_3-2\right) \left(2 t_1-1\right) t_2\nonumber\\
&&\hspace{-0.2cm}
+3 s_1+1+\frac{2s_1 s_4t_3^2}{\left(s_2+s_3-1\right)^2}
+\frac{s_1t_3 \left(-2 s_4 t_3-2s_1+1\right)}{s_2+s_3-1}\nonumber\\
&&\hspace{-0.2cm}
+\frac{x_i}{2}\Bigl(2 (t_2-t_3)+3 (t_3-1) (2 s_1+s_2+s_3)\Bigr),
\end{eqnarray}
\begin{eqnarray}
N_{2}^{(a1)}+N_{2}^{(a2)}
\hspace{-0.2cm}&=&\hspace{-0.2cm}
-20 s_2 t_2^2-2 \left(s_1-s_2-12\right) t_2-4 \left(5
		 \left(t_3+t_4\right)-4\right) \left(s_1 t_1+s_2
		 t_2\right)\nonumber\\
&&\hspace{-0.2cm}
+2t_1-8+\frac{s_1 \left(-8 t_2+4 \left(10
   \left(1-t_1\right)-23\right) \left(t_3+t_4\right)+19\right)}{2
		 \left(s_2-1\right)}+\frac{17 s_1-1}{2 s_2}\nonumber\\
&&\hspace{-0.2cm}
+\frac{x_i}{2}\frac{s_2\Bigl(-6 t_2 (s_2-1)+s_2-1\Bigr)+s_1\Bigl(s_2 (6 t_3+6
t_4-5)+4\Bigr)}{(s_2-1) s_2}
\nonumber\\
&&\hspace{-0.2cm}
+\frac{x_j}{2}\frac{s_1\Bigl(5-6 s_2 (t_1+t_2)\Bigr)-(s_2-1) (6 t_2
s_2-1)}{s_2-1},
\end{eqnarray}
%%%
\begin{eqnarray}
N_{2}^{(a3)}
\hspace{-0.2cm}&=&\hspace{-0.2cm}
-\frac{2 s_4^2 \left(4 \left(t_3+t_4\right)^2-6
		\left(t_3+t_4\right)+1\right)}{-s_1-s_4}
- 8s_4t_3^2-2s_4t_3\left(8 t_4-7\right)\nonumber\\
&&\hspace{-0.2cm}
-s_4\left(t_4-1\right)\left(8 t_4-3\right)
-8 \left(s_1+s_4\right) t_2^2+3 s_1 t_3+3
   \left(s_2+s_3\right) \left(1-t_4\right)\nonumber\\
&&\hspace{-0.2cm}
+\frac{2 \left(s_4-2\right)
   \left(s_4
   \left(t_3-2\right)+1\right)-3 s_1 \left(2
   t_2+t_3\right)+\left(s_4-3\right) \left(2 s_4-1\right) t_4+9
   t_2}{s_2+s_3}\nonumber\\
&&\hspace{-0.2cm}
+t_2 \Bigl(-8 \left(s_1+s_4\right)
   \left(t_3+t_4\right)+6 s_1-7 \left(s_2+s_3\right)-2\Bigr)+1\nonumber\\
&&\hspace{-0.2cm}
+\frac{x_i}{2}\Bigl(3 (t_3+t_4+1) (2s_4+s_3+s_2)-3 (t_3+t_4)-1\Bigr).
\end{eqnarray}
Note that since the diagrams (a1) and (a2) give the same contribution, one
has  the relation $N_{n}^{(a1)}=N_{n}^{(a2)}$. 
The numerators for the diagrams (b1) and (b2) are given by
\begin{eqnarray}
N_{1}^{(b1)}
\hspace{-0.2cm}&=&\hspace{-0.2cm}
\frac{3 s_1 \left(t_2+t_3\right) \Bigl(\left(s_3-s_2\right)
	     \left(1+s_4+s_5\right)
	     \left(t_2+t_3\right)+\left(2-s_2+s_3\right)
   \left(s_1+s_2+s_3\right)\Bigr)}{2 \left(s_4+s_5-1\right){}^3
   \left(s_4+s_5\right)}\nonumber\\
&&\hspace{-0.2cm}
+\frac{x_i}{2}\frac{\left(2 s_3+s_4+s_5-1\right) \left(t_2+t_3\right)}
{\left(s_4+s_5-1\right)^2\left(s_4+s_5\right)},
\end{eqnarray}
%%%
\begin{eqnarray}
N_{1}^{(b2)}
\hspace{-0.2cm}&=&\hspace{-0.2cm}
\frac{\left(s_3-s_2\right) \Bigl(3 \left(s_1+s_2+s_3\right)
       t_1-t_1+1\Bigr)}{\left(s_4+s_5-1\right)^2
       \left(s_4+s_5\right)}-\frac{2 s_1
   \left(s_2-s_3\right)
   \left(1-t_1\right)^2}{\left(s_4+s_5-1\right)^3
   \left(s_4+s_5\right)}\nonumber\\
&&\hspace{-0.2cm}
+\frac{x_i}{2}\frac{2 \left(s_2-s_3\right)
		 \left(1-t_1\right)+\left(s_1+s_2-s_3\right) \left(4-3s_4-3s_5\right)
		 \left(t_1-2\right)}{
   \left(s_4+s_5-1\right){}^2 \left(s_4+s_5\right)},
\end{eqnarray}

\begin{eqnarray}
N_{2}^{(b1)}
\hspace{-0.2cm}&=&\hspace{-0.2cm}
-\frac{6 (2 s_3-(s_4+s_5) ((1-3 t_1) s_1-t_1 (s_2-s_3)+1))}
{(s_4+s_5-1)^2 (s_4+s_5)^2}\nonumber\\
&&\hspace{-0.2cm}
+\frac{s_1t_1(s_2-s_3)(20t_1-3)}
{(s_4+s_5-1)^3}
+\frac{s_1 (s_2-s_3) \left(6 (1-5t_1)(s_4+s_5)+7\right)}
{(s_4+s_5-1)^3 (s_4+s_5)^2}\nonumber\\
&&\hspace{-0.2cm}
+\frac{3 (s_2-s_3) (t_1 (s_4+s_5)-1)}{(s_4+s_5-1)^3 (s_4+s_5)}\nonumber\\
&&\hspace{-0.2cm}
-x_i\frac{2s_1(5-3t_1)(s_4+s_5) +
s_1(1-s_1-2s_3)+(s_2-s_3+1)(s_2+s_3)}
{(s_4+s_5-1)^2(s_4+s_5)^2},
\end{eqnarray}
%%%
\begin{eqnarray}
N_{2}^{(b2)}
\hspace{-0.2cm}&=&\hspace{-0.2cm}
-\frac{s_1 (t_2+t_3)^2 (s_2-s_3)}{(s_4+s_5-1)^3 (s_4+s_5)}
+\frac{2 (s_2-s_3) (1-3 t_1 (s_4+s_5))}{(s_4+s_5-1) (s_4+s_5)^2}\nonumber\\
&&\hspace{-0.2cm}
+\frac{2 s_1 (s_2-s_3)}{(s_4+s_5-1)^2 (s_4+s_5)^2}\nonumber\\
&&\hspace{-0.2cm}
-x_i\left(
\frac{3t_1}{s_4+s_5}-\frac{3 t_1 s_5 (s_1+s_2-1)-3 t_1 s_3
s_4 + 2(s_1+s_3-1)}{(s_4+s_5-1) (s_4+s_5)^2}\right).
\end{eqnarray}

%%%%%%%%%%%%%%%%%%%%%%%%%%%%%%%%%%%


\begin{thebibliography}{200}
%\bibliographystyle{unsrt}

%\cite{Aad:2012tfa}
%\cite{Aad:2012tfa}
\bibitem{Aad:2012tfa}
  G.~Aad {\it et al.} [ATLAS Collaboration],
  %``Observation of a new particle in the search for the Standard Model Higgs boson with the ATLAS detector at the LHC,''
  Phys.\ Lett.\ B {\bf 716} (2012) 1
  [arXiv:1207.7214 [hep-ex]].
  %%CITATION = ARXIV:1207.7214;%%
  %5133 citations counted in INSPIRE as of 07 Nov 2015

%\cite{Chatrchyan:2012xdj}
\bibitem{Chatrchyan:2012xdj}
  S.~Chatrchyan {\it et al.} [CMS Collaboration],
  %``Observation of a new boson at a mass of 125 GeV with the CMS experiment at the LHC,''
  Phys.\ Lett.\ B {\bf 716} (2012) 30
  [arXiv:1207.7235 [hep-ex]].
  %%CITATION = ARXIV:1207.7235;%%
  %5035 citations counted in INSPIRE as of 07 Nov 2015
  
  %\cite{Minkowski:1977sc}
\bibitem{Minkowski:1977sc}
  P.~Minkowski,
  %``mu --> e gamma at a Rate of One Out of 1-Billion Muon Decays?,''
  Phys.\ Lett.\ B {\bf 67} (1977) 421.
  %%CITATION = PHLTA,B67,421;%%

%\cite{Yanagida:1979as}
\bibitem{Yanagida:1979as}
  T.~Yanagida,
  %``Horizontal Symmetry And Masses Of Neutrinos,''
  Conf.\ Proc.\ C {\bf 7902131} (1979) 95
   [Conf.\ Proc.\ C {\bf 7902131} (1979) 95].
  %%CITATION = CONFP,C7902131,95;%%
  %1091 citations counted in INSPIRE as of 07 Nov 2015
  
%\cite{GellMann:1980vs}
\bibitem{GellMann:1980vs} 
  M.~Gell-Mann, P.~Ramond and R.~Slansky,
  %``Complex Spinors and Unified Theories,''
  Conf.\ Proc.\ C {\bf 790927} (1979) 315 
  [arXiv:1306.4669 [hep-th]].
  %%CITATION = ARXIV:1306.4669;%%
  %2152 citations counted in INSPIRE as of 07 sept. 2015

%\cite{Glashow:1979nm}
\bibitem{Glashow:1979nm}
  S.~L.~Glashow,
  %``The Future of Elementary Particle Physics,''
  NATO Sci.\ Ser.\ B {\bf 61} (1980) 687.
  %318 citations counted in INSPIRE as of 16 Oct 2015
  
%\cite{Mohapatra:1979ia}
\bibitem{Mohapatra:1979ia}
  R.~N.~Mohapatra and G.~Senjanovic,
  %``Neutrino Mass and Spontaneous Parity Violation,''
  Phys.\ Rev.\ Lett.\  {\bf 44} (1980) 912.
  %%CITATION = PRLTA,44,912;%%

%\cite{Schechter:1980gr}
\bibitem{Schechter:1980gr}
  J.~Schechter and J.~W.~F.~Valle,
  %``Neutrino Masses In SU(2) X U(1) Theories,''
  Phys.\ Rev.\ D {\bf 22} (1980) 2227.
  %%CITATION = PHRVA,D22,2227;%%

%\cite{Schechter:1981cv}
\bibitem{Schechter:1981cv}
  J.~Schechter and J.~W.~F.~Valle,
  %``Neutrino Decay and Spontaneous Violation of Lepton Number,''
  Phys.\ Rev.\ D {\bf 25} (1982) 774.
  %%CITATION = PHRVA,D25,774;%%

  %\cite{Asaka:2005an}
\bibitem{Asaka:2005an} 
  T.~Asaka, S.~Blanchet and M.~Shaposhnikov,
  %``The nuMSM, dark matter and neutrino masses,''
  Phys.\ Lett.\ B {\bf 631} (2005) 151 
  [hep-ph/0503065].
  %%CITATION = HEP-PH/0503065;%%
  %354 citations counted in INSPIRE as of 05 Oct 2015


%\cite{Mohapatra:1986bd}
\bibitem{Mohapatra:1986bd} 
  R.~N.~Mohapatra and J.~W.~F.~Valle,
  %``Neutrino Mass and Baryon Number Nonconservation in Superstring
	%Models,''
  Phys.\ Rev.\ D {\bf 34} 1642 (1986).
  %%CITATION = PHRVA,D34,1642;%%
  %664 citations counted in INSPIRE as of 07 sept. 2015

%\cite{GonzalezGarcia:1988rw}
\bibitem{GonzalezGarcia:1988rw}
  M.~C.~Gonzalez-Garcia and J.~W.~F.~Valle,
  %``Fast Decaying Neutrinos and Observable Flavor Violation in a New Class of Majoron Models,''
  Phys.\ Lett.\ B {\bf 216} (1989) 360.
  %%CITATION = PHLTA,B216,360;%%
  %216 citations counted in INSPIRE as of 16 Oct 2015

%\cite{Deppisch:2004fa}
\bibitem{Deppisch:2004fa}
  F.~Deppisch and J.~W.~F.~Valle,
  %``Enhanced lepton flavor violation in the supersymmetric inverse seesaw model,''
  Phys.\ Rev.\ D {\bf 72} (2005) 036001
  [hep-ph/0406040].
  %%CITATION = HEP-PH/0406040;%%
  %137 citations counted in INSPIRE as of 16 Oct 2015

%\cite{Gavela:2009cd}
\bibitem{Gavela:2009cd}
  M.~B.~Gavela, T.~Hambye, D.~Hernandez and P.~Hernandez,
  %``Minimal Flavour Seesaw Models,''
  JHEP {\bf 0909} (2009) 038
  [arXiv:0906.1461 [hep-ph]].
  %%CITATION = ARXIV:0906.1461;%%
  %116 citations counted in INSPIRE as of 16 Oct 2015
  
%\cite{Ibarra:2010xw}
\bibitem{Ibarra:2010xw}
  A.~Ibarra, E.~Molinaro and S.~T.~Petcov,
  %``TeV Scale See-Saw Mechanisms of Neutrino Mass Generation, the Majorana Nature of the Heavy Singlet Neutrinos and $(\beta\beta)_{0\nu}$-Decay,''
  JHEP {\bf 1009} (2010) 108
  [arXiv:1007.2378 [hep-ph]].
  %%CITATION = ARXIV:1007.2378;%%
  %101 citations counted in INSPIRE as of 16 Oct 2015

%\cite{Abada:2014vea}
\bibitem{Abada:2014vea}
  A.~Abada and M.~Lucente,
  %``Looking for the minimal inverse seesaw realisation,''
  Nucl.\ Phys.\ B {\bf 885} (2014) 651
  [arXiv:1401.1507 [hep-ph]].
  %%CITATION = ARXIV:1401.1507;%%
  %15 citations counted in INSPIRE as of 07 Nov 2015}
  
%\cite{Barr:2003nn}
\bibitem{Barr:2003nn}
  S.~M.~Barr,
  %``A Different seesaw formula for neutrino masses,''
  Phys.\ Rev.\ Lett.\  {\bf 92} (2004) 101601
  [hep-ph/0309152].
  %%CITATION = HEP-PH/0309152;%%
  %78 citations counted in INSPIRE as of 16 Oct 2015

%\cite{Malinsky:2005bi}
\bibitem{Malinsky:2005bi} 
  M.~Malinsky, J.~C.~Romao and J.~W.~F.~Valle,
  %``Novel supersymmetric SO(10) seesaw mechanism,''
  Phys.\ Rev.\ Lett.\  {\bf 95} (2005) 161801 
  [hep-ph/0506296].
  %%CITATION = HEP-PH/0506296;%%
  %136 citations counted in INSPIRE as of 01 Oct 2015

%\cite{Abada:2014zra}
\bibitem{Abada:2014zra}
  A.~Abada, G.~Arcadi and M.~Lucente,
  %``Dark Matter in the minimal Inverse Seesaw mechanism,''
  JCAP {\bf 1410} (2014) 001
  [arXiv:1406.6556 [hep-ph]].
  %%CITATION = ARXIV:1406.6556;%%
  %24 citations counted in INSPIRE as of 16 Oct 2015

%\cite{Akhmedov:1998qx}
\bibitem{Akhmedov:1998qx} 
  E.~K.~Akhmedov, V.~A.~Rubakov and A.~Y.~Smirnov,
  %``Baryogenesis via neutrino oscillations,''
  Phys.\ Rev.\ Lett.\  {\bf 81} (1998) 1359 
  [hep-ph/9803255].
  %%CITATION = HEP-PH/9803255;%%
  %207 citations counted in INSPIRE as of 05 Oct 2015


%\cite{Canetti:2012vf}
\bibitem{Canetti:2012vf}
  L.~Canetti, M.~Drewes and M.~Shaposhnikov,
  %``Sterile Neutrinos as the Origin of Dark and Baryonic Matter,''
  Phys.\ Rev.\ Lett.\  {\bf 110} (2013) 6,  061801
  %doi:10.1103/PhysRevLett.110.061801
  [arXiv:1204.3902 [hep-ph]].
  %%CITATION = doi:10.1103/PhysRevLett.110.061801;%%
  %81 citations counted in INSPIRE as of 15 Jan 2016
  
%\cite{Canetti:2012kh}
\bibitem{Canetti:2012kh}
  L.~Canetti, M.~Drewes, T.~Frossard and M.~Shaposhnikov,
  %``Dark Matter, Baryogenesis and Neutrino Oscillations from Right Handed Neutrinos,''
  Phys.\ Rev.\ D {\bf 87} (2013) 093006
  %doi:10.1103/PhysRevD.87.093006
  [arXiv:1208.4607 [hep-ph]].
  %%CITATION = doi:10.1103/PhysRevD.87.093006;%%
  %114 citations counted in INSPIRE as of 15 Jan 2016

%\cite{Abada:2015rta}
\bibitem{Abada:2015rta} 
  A.~Abada, G.~Arcadi, V.~Domcke and M.~Lucente,
  %``Lepton number violation as a key to low-scale leptogenesis,''
  JCAP {\bf 1511}, no. 11, 041 (2015)
  %doi:10.1088/1475-7516/2015/11/041
  [arXiv:1507.06215 [hep-ph]].
  %%CITATION = doi:10.1088/1475-7516/2015/11/041;%%
  %6 citations counted in INSPIRE as of 13 Jan 2016

%\cite{Canetti:2014dka}
\bibitem{Canetti:2014dka}
  L.~Canetti, M.~Drewes and B.~Garbrecht,
  %``Probing leptogenesis with GeV-scale sterile neutrinos at LHCb and Belle II,''
  Phys.\ Rev.\ D {\bf 90} (2014) 12,  125005
  %doi:10.1103/PhysRevD.90.125005
  [arXiv:1404.7114 [hep-ph]].
  %%CITATION = doi:10.1103/PhysRevD.90.125005;%%
  %23 citations counted in INSPIRE as of 15 Jan 2016


%\cite{Mueller:2011nm}
\bibitem{Mueller:2011nm}
  T.~A.~Mueller, D.~Lhuillier, M.~Fallot, A.~Letourneau, S.~Cormon,
  M.~Fechner, L.~Giot and T.~Lasserre {\it et al.},
  %``Improved Predictions of Reactor Antineutrino Spectra,''
  Phys.\ Rev.\ C {\bf 83} (2011) 054615
  [arXiv:1101.2663 [hep-ex]].

%\cite{Huber:2011wv}
\bibitem{Huber:2011wv}
  P.~Huber,
  %``On the determination of anti-neutrino spectra from nuclear reactors,''
  Phys.\ Rev.\ C {\bf 84} (2011) 024617
   [Erratum-ibid.\ C {\bf 85} (2012) 029901]
  [arXiv:1106.0687 [hep-ph]].

  %\cite{Mention:2011rk}
\bibitem{Mention:2011rk}
  G.~Mention, M.~Fechner, T.~Lasserre, T.~A.~Mueller, D.~Lhuillier, M.~Cribier and A.~Letourneau,
  %``The Reactor Antineutrino Anomaly,''
  Phys.\ Rev.\ D {\bf 83} (2011) 073006
  [arXiv:1101.2755 [hep-ex]].

%\cite{Aguilar:2001ty}
\bibitem{Aguilar:2001ty}
  A.~Aguilar-Arevalo {\it et al.} [LSND Collaboration],
  %``Evidence for neutrino oscillations from the observation of
  %anti-neutrino(electron) appearance in a anti-neutrino(muon)
  %beam,'' 
  Phys.\ Rev.\ D {\bf 64} (2001) 112007
  [hep-ex/0104049].
  %%CITATION = HEP-EX/0104049;%%
  %1323 citations counted in INSPIRE as of 16 Oct 2015

%\cite{AguilarArevalo:2007it}
\bibitem{AguilarArevalo:2007it}
  A.~A.~Aguilar-Arevalo {\it et al.}  [MiniBooNE Collaboration],
  %``A Search for electron neutrino appearance at the $\Delta m^{2} \sim 1$eV$^{2}$ scale,''
  Phys.\ Rev.\ Lett.\  {\bf 98} (2007) 231801
  [arXiv:0704.1500 [hep-ex]].

  %\cite{AguilarArevalo:2010wv}
\bibitem{AguilarArevalo:2010wv}
  A.~A.~Aguilar-Arevalo {\it et al.}  [MiniBooNE Collaboration],
  %``Event Excess in the MiniBooNE Search for $\bar \nu_\mu \rightarrow \bar \nu_e$ Oscillations,''
  Phys.\ Rev.\ Lett.\  {\bf 105} (2010) 181801
  [arXiv:1007.1150 [hep-ex]].

  %\cite{Aguilar-Arevalo:2013pmq}
\bibitem{Aguilar-Arevalo:2013pmq}
  A.~A.~Aguilar-Arevalo {\it et al.}  [MiniBooNE Collaboration],
  %``Improved Search for $\bar \nu_\mu \rightarrow \bar \nu_e$ Oscillations in the MiniBooNE Experiment,''
  Phys.\ Rev.\ Lett.\  {\bf 110} (2013) 161801
  [arXiv:1207.4809 [hep-ex], arXiv:1303.2588 [hep-ex]].
  
  %\cite{Acero:2007su}
\bibitem{Acero:2007su}
  M.~A.~Acero, C.~Giunti and M.~Laveder,
  %``Limits on nu(e) and anti-nu(e) disappearance from Gallium and reactor experiments,''
  Phys.\ Rev.\ D {\bf 78} (2008) 073009
  [arXiv:0711.4222 [hep-ph]].

  %\cite{Giunti:2010zu}
\bibitem{Giunti:2010zu}
  C.~Giunti and M.~Laveder,
  %``Statistical Significance of the Gallium Anomaly,''
  Phys.\ Rev.\ C {\bf 83} (2011) 065504
  [arXiv:1006.3244 [hep-ph]].

  
   \bibitem{Kusenko:2009up}
  A.~Kusenko,
%  \emph{Sterile neutrinos: The Dark side of the light fermions},
  Phys.\ Rept.\  {\bf 481} (2009) 1
  [arXiv:0906.2968 [hep-ph]].
  
   \bibitem{Abazajian:2012ys}
  K.~N.~Abazajian, M.~A.~Acero, S.~K.~Agarwalla, A.~A.~Aguilar-Arevalo,
	   C.~H.~Albright, S.~Antusch, C.~A.~Arguelles and
	   A.~B.~Balantekin {\it et al.}, 
  %\emph{Light Sterile Neutrinos: A White Paper},
  arXiv:1204.5379 [hep-ph].
  %%CITATION = ARXIV:1204.5379;%%

%\cite{Dev:2013wba}
\bibitem{Dev:2013wba} 
  P.~S.~B.~Dev, A.~Pilaftsis and U.~k.~Yang,
  %``New Production Mechanism for Heavy Neutrinos at the LHC,''
  Phys.\ Rev.\ Lett.\  {\bf 112} (2014) 081801 
  [arXiv:1308.2209 [hep-ph]].
  %%CITATION = ARXIV:1308.2209;%%
  %36 citations counted in INSPIRE as of 02 Oct 2015


%\cite{Das:2014jxa}
\bibitem{Das:2014jxa}
  A.~Das, P.~S.~Bhupal Dev and N.~Okada,
  %``Direct bounds on electroweak scale pseudo-Dirac neutrinos from $\sqrt s=8$ TeV LHC data,''
  Phys.\ Lett.\ B {\bf 735} (2014) 364
  %doi:10.1016/j.physletb.2014.06.058
  [arXiv:1405.0177 [hep-ph]].


%\cite{Das:2015toa}
\bibitem{Das:2015toa}
  A.~Das and N.~Okada,
  %``Improved bounds on the heavy neutrino productions at the LHC,''
  arXiv:1510.04790 [hep-ph].
 

  \bibitem{Adams:2013qkq}
  C.~Adams {\it et al.} [LBNE Collaboration],
  %``The Long-Baseline Neutrino Experiment: Exploring Fundamental Symmetries of the Universe,'' 
  arXiv:1307.7335 [hep-ex].

 \bibitem{Bonivento:2013jag}
  W.~Bonivento {\it et al.},
  %``Proposal to Search for Heavy Neutral Leptons at the SPS,''
  arXiv:1310.1762 [hep-ex].


\bibitem{Blondel:2013ia} 
  A.~Blondel {\it et al.},
  %``Research Proposal for an Experiment to Search for the Decay $\mu
	%\to eee$,''
  arXiv:1301.6113 [physics.ins-det].
  %%CITATION = ARXIV:1301.6113;%%
  %65 citations counted in INSPIRE as of 02 Oct 2015

%\cite{Deppisch:2015qwa}
\bibitem{Deppisch:2015qwa} 
  F.~F.~Deppisch, P.~S.~Bhupal Dev and A.~Pilaftsis,
  %``Neutrinos and Collider Physics,''
  New J.\ Phys.\  {\bf 17} (2015) 075019 
  [arXiv:1502.06541 [hep-ph]].

%\cite{Atre:2009rg}
\bibitem{Atre:2009rg} 
  A.~Atre, T.~Han, S.~Pascoli and B.~Zhang,
  %``The Search for Heavy Majorana Neutrinos,''
  JHEP {\bf 0905} (2009) 030 
  [arXiv:0901.3589 [hep-ph]].
  %%CITATION = ARXIV:0901.3589;%%
  %247 citations counted in INSPIRE as of 19 Jul 2015

%\cite{Alekhin:2015byh}
\bibitem{Alekhin:2015byh} 
  S.~Alekhin {\it et al.},
  %``A facility to Search for Hidden Particles at the CERN SPS: the SHiP
	%physics case,''
  arXiv:1504.04855 [hep-ph].
 
  \bibitem{Anelli:2015pba}
  M.~Anelli {\it et al.} [SHiP Collaboration],
  %``A facility to Search for Hidden Particles (SHiP) at the CERN SPS,''
  arXiv:1504.04956 [physics.ins-det].

%\cite{Banerjee:2015gca}
\bibitem{Banerjee:2015gca} 
  S.~Banerjee, P.~S.~B.~Dev, A.~Ibarra, T.~Mandal and M.~Mitra,
  %``Prospects of Heavy Neutrino Searches at Future Lepton Colliders,''
  Phys.\ Rev.\ D {\bf 92} (2015) 075002 
  [arXiv:1503.05491 [hep-ph]].

%\cite{Das:2012ze}
\bibitem{Das:2012ze}
  A.~Das and N.~Okada,
  %``Inverse seesaw neutrino signatures at the LHC and ILC,''
  Phys.\ Rev.\ D {\bf 88} (2013) 113001
  %doi:10.1103/PhysRevD.88.113001
  [arXiv:1207.3734 [hep-ph]].


%\cite{deGouvea:2015euy}
\bibitem{deGouvea:2015euy}
  A.~de Gouv\^ea and A.~Kobach,
  %``Global Constraints on a Heavy Neutrino,''
  arXiv:1511.00683 [hep-ph].
  %%CITATION = ARXIV:1511.00683;%%
  




%\cite{Ng:1995cs}
\bibitem{Ng:1995cs} 
  D.~Ng and J.~N.~Ng,
  %``A Note on Majorana neutrinos, leptonic CKM and electron electric
	%dipole moment,''
  Mod.\ Phys.\ Lett.\ A {\bf 11}  (1996) 211
  [hep-ph/9510306].
  %%CITATION = HEP-PH/9510306;%%
  %18 citations counted in INSPIRE as of 29 Aug 2015

%\cite{Archambault:2004td}
\bibitem{Archambault:2004td} 
  J.~P.~Archambault, A.~Czarnecki and M.~Pospelov,
  %``Electric dipole moments of leptons in the presence of majorana
	%neutrinos,''
  Phys.\ Rev.\ D {\bf 70} (2004) 073006 
  [hep-ph/0406089].
  %%CITATION = HEP-PH/0406089;%%
  %16 citations counted in INSPIRE as of 29 Aug 2015

%\cite{Chang:2004pba}
\bibitem{Chang:2004pba} 
  W.~F.~Chang and J.~N.~Ng,
  %``Charged lepton electric dipole moments from TeV scale right-handed
	%neutrinos,''
  New J.\ Phys.\  {\bf 7} (2005) 65 
  [hep-ph/0411201].
  %%CITATION = HEP-PH/0411201;%%
  %9 citations counted in INSPIRE as of 29 Aug 2015


%\cite{Ellis:2001yza}
\bibitem{Ellis:2001yza}
  J.~R.~Ellis, J.~Hisano, M.~Raidal and Y.~Shimizu,
  %``Lepton electric dipole moments in nondegenerate supersymmetric seesaw models,''
  Phys.\ Lett.\ B {\bf 528} (2002) 86
  %doi:10.1016/S0370-2693(02)01197-8
  [hep-ph/0111324].
  %%CITATION = doi:10.1016/S0370-2693(02)01197-8;%%
  %91 citations counted in INSPIRE as of 15 Jan 2016
  
%\cite{Masina:2003wt}
\bibitem{Masina:2003wt}
  I.~Masina,
  %``Lepton electric dipole moments from heavy states Yukawa couplings,''
  Nucl.\ Phys.\ B {\bf 671} (2003) 432
  %doi:10.1016/j.nuclphysb.2003.08.018
  [hep-ph/0304299].
  %%CITATION = doi:10.1016/j.nuclphysb.2003.08.018;%%
  %44 citations counted in INSPIRE as of 15 Jan 2016
  
%\cite{Farzan:2004qu}
\bibitem{Farzan:2004qu}
  Y.~Farzan and M.~E.~Peskin,
  %``The Contribution from neutrino Yukawa couplings to lepton electric dipole moments,''
  Phys.\ Rev.\ D {\bf 70} (2004) 095001
  %doi:10.1103/PhysRevD.70.095001
  [hep-ph/0405214].
  %%CITATION = doi:10.1103/PhysRevD.70.095001;%%
  %39 citations counted in INSPIRE as of 15 Jan 2016



%\cite{Ilakovac:1994kj}
\bibitem{Ilakovac:1994kj} 
  A.~Ilakovac and A.~Pilaftsis,
  %``Flavor violating charged lepton decays in seesaw-type models,''
  Nucl.\ Phys.\ B {\bf 437} (1995) 491 
  [hep-ph/9403398].
  %%CITATION = HEP-PH/9403398;%%
  %244 citations counted in INSPIRE as of 18 Aug 2015

%\cite{Alonso:2012ji}
\bibitem{Alonso:2012ji} 
  R.~Alonso, M.~Dhen, M.~B.~Gavela and T.~Hambye,
  %``Muon conversion to electron in nuclei in type-I seesaw models,''
  JHEP {\bf 1301} (2013) 118 
  [arXiv:1209.2679 [hep-ph]].
  %%CITATION = ARXIV:1209.2679;%%
  %47 citations counted in INSPIRE as of 13 Aug 2015

%\cite{Gonzalez-Garcia:2014bfa}
\bibitem{Gonzalez-Garcia:2014bfa} 
  M.~C.~Gonzalez-Garcia, M.~Maltoni and T.~Schwetz,
  %``Updated fit to three neutrino mixing: status of leptonic CP
	%violation,''
  JHEP {\bf 1411} (2014) 052 
  [arXiv:1409.5439 [hep-ph]].

%\cite{Pospelov:2005pr}
\bibitem{Pospelov:2005pr} 
  M.~Pospelov and A.~Ritz,
  %``Electric dipole moments as probes of new physics,''
  Annals Phys.\  {\bf 318} (2005) 119 
  [hep-ph/0504231].
  %%CITATION = HEP-PH/0504231;%%
  %399 citations counted in INSPIRE as of 12 Oct 2015

%\cite{Fukuyama:2012np}
\bibitem{Fukuyama:2012np} 
  T.~Fukuyama,
  %``Searching for New Physics beyond the Standard Model in Electric
	%Dipole Moment,''
  Int.\ J.\ Mod.\ Phys.\ A {\bf 27} (2012) 1230015 
  [arXiv:1201.4252 [hep-ph]].
  %%CITATION = ARXIV:1201.4252;%%
  %41 citations counted in INSPIRE as of 13 Jul 2015

%\cite{Agashe:2014kda}
\bibitem{Agashe:2014kda} 
  K.~A.~Olive {\it et al.} [Particle Data Group Collaboration],
  %``Review of Particle Physics,''
  Chin.\ Phys.\ C {\bf 38} (2014) 090001.
  %%CITATION = CHPHD,C38,090001;%%
  %1686 citations counted in INSPIRE as of 18 Aug 2015

%\cite{Baron:2013eja}
\bibitem{Baron:2013eja} 
  J.~Baron {\it et al.}  [ACME Collaboration],
  %``Order of Magnitude Smaller Limit on the Electric Dipole Moment of
	%the Electron,''
  Science {\bf 343} (2014) 269 
  [arXiv:1310.7534 [physics.atom-ph]].
  %%CITATION = ARXIV:1310.7534;%%
  %132 citations counted in INSPIRE as of 26 May 2015

%\cite{Bennett:2008dy}
\bibitem{Bennett:2008dy} 
  G.~W.~Bennett {\it et al.} [Muon (g-2) Collaboration],
  %``An Improved Limit on the Muon Electric Dipole Moment,''
  Phys.\ Rev.\ D {\bf 80}  (2009) 052008 
  [arXiv:0811.1207 [hep-ex]].
  %%CITATION = ARXIV:0811.1207;%%
  %92 citations counted in INSPIRE as of 13 Jul 2015

%\cite{Inami:2002ah}
\bibitem{Inami:2002ah} 
  K.~Inami {\it et al.} [Belle Collaboration],
  %``Search for the electric dipole moment of the tau lepton,''
  Phys.\ Lett.\ B {\bf 551} (2003) 16 
  [hep-ex/0210066].
  %%CITATION = HEP-EX/0210066;%%

%\cite{Greenberg:2002uu}
\bibitem{Greenberg:2002uu} 
  O.~W.~Greenberg,
  %``CPT violation implies violation of Lorentz invariance,''
  Phys.\ Rev.\ Lett.\  {\bf 89} (2002) 231602 
  [hep-ph/0201258].
  %%CITATION = HEP-PH/0201258;%%
  %246 citations counted in INSPIRE as of 10 sept. 2015

\bibitem{acme:next_generation}
  W.~C.~Griffith, 
  %``Measurements and implications of of EDMs''
  Plenary talk at ``Interplay between Particle \& Astroparticle
  physics 2014'', 
  \url{https://indico.ph.qmul.ac.uk/indico/conferenceDisplay.py?confId=1}. 

%\cite{Saito:2012zz}
\bibitem{Saito:2012zz}
  N.~Saito [J-PARC g-'2/EDM Collaboration],
  %``A novel precision measurement of muon g-2 and EDM at J-PARC,''
  AIP Conf.\ Proc.\  {\bf 1467} (2012) 45.
  %%CITATION = APCPC,1467,45;%%
  %20 citations counted in INSPIRE as of 07 Nov 2015

%\cite{Shabalin:1978rs}
\bibitem{Shabalin:1978rs} 
  E.~P.~Shabalin,
  %``Electric Dipole Moment of Quark in a Gauge Theory with Left-Handed
	%Currents,''
  Sov.\ J.\ Nucl.\ Phys.\  {\bf 28}, 75 (1978)
  [Yad.\ Fiz.\  {\bf 28} 151 (1978)].
  %%CITATION = SJNCA,28,75;%%
  %191 citations counted in INSPIRE as of 10 sept. 2015

%\cite{Pilaftsis:1997dr}
\bibitem{Pilaftsis:1997dr} 
  A.~Pilaftsis,
  %``Resonant CP violation induced by particle mixing in transition
	%amplitudes,''
  Nucl.\ Phys.\ B {\bf 504} (1997) 61 
  [hep-ph/9702393].
  %%CITATION = HEP-PH/9702393;%%
  %185 citations counted in INSPIRE as of 12 Oct 2015

%\cite{Buchmuller:1997yu}
\bibitem{Buchmuller:1997yu} 
  W.~Buchmuller and M.~Plumacher,
  %``CP asymmetry in Majorana neutrino decays,''
  Phys.\ Lett.\ B {\bf 431} (1998) 354 
  [hep-ph/9710460].
  %%CITATION = HEP-PH/9710460;%%
  %278 citations counted in INSPIRE as of 12 Oct 2015

%\cite{Mertig:1990an}
\bibitem{Mertig:1990an} 
  R.~Mertig, M.~Bohm and A.~Denner,
  %``FEYN CALC: Computer algebraic calculation of Feynman amplitudes,''
  Comput.\ Phys.\ Commun.\  {\bf 64} (1991) 345.
  %%CITATION = CPHCB,64,345;%%
  %499 citations counted in INSPIRE as of 12 Aug 2015

%\cite{Abada:2015zea}
\bibitem{Abada:2015zea} 
  A.~Abada, D.~Be\v{c}irevi\'c, M.~Lucente and O.~Sumensari,
  %``Lepton flavor violating decays of vector quarkonia and of the $Z$
	%boson,''
  Phys.\ Rev.\ D {\bf 91}, (2015) 113013 
  [arXiv:1503.04159 [hep-ph]].
  %%CITATION = ARXIV:1503.04159;%%
  %1 citations counted in INSPIRE as of 19 Jul 2015

%\cite{Adam:2013mnn}
\bibitem{Adam:2013mnn} 
  J.~Adam {\it et al.} [MEG Collaboration],
  %``New constraint on the existence of the $\mu^+ \to e^+\gamma$
	%decay,''
  Phys.\ Rev.\ Lett.\  {\bf 110}  (2013) 201801
  [arXiv:1303.0754 [hep-ex]].
  %%CITATION = ARXIV:1303.0754;%%
  %304 citations counted in INSPIRE as of 02 Oct 2015

%\cite{Baldini:2013ke}
\bibitem{Baldini:2013ke} 
  A.~M.~Baldini {\it et al.},
  %``MEG Upgrade Proposal,''
  arXiv:1301.7225 [physics.ins-det].
  %%CITATION = ARXIV:1301.7225;%%
  %105 citations counted in INSPIRE as of 02 Oct 2015

%\cite{Bellgardt:1987du}
\bibitem{Bellgardt:1987du} 
  U.~Bellgardt {\it et al.} [SINDRUM Collaboration],
  %``Search for the Decay mu+ ---> e+ e+ e-,''
  Nucl.\ Phys.\ B {\bf 299} (1988) 1.
  %%CITATION = NUPHA,B299,1;%%
  %415 citations counted in INSPIRE as of 02 Oct 2015

%\cite{Abreu:1996pa}
\bibitem{Abreu:1996pa} 
  P.~Abreu {\it et al.} [DELPHI Collaboration],
  %``Search for neutral heavy leptons produced in Z decays,''
  Z.\ Phys.\ C {\bf 74} (1997) 57 
  [Z.\ Phys.\ C {\bf 75} (1997) 580 ].
  %%CITATION = ZEPYA,C74,57;%%
  %94 citations counted in INSPIRE as of 07 sept. 2015

%\cite{Adriani:1992pq}
\bibitem{Adriani:1992pq} 
  O.~Adriani {\it et al.} [L3 Collaboration],
  %``Search for isosinglet neutral heavy leptons in Z0 decays,''
  Phys.\ Lett.\ B {\bf 295} (1992) 371.
  %%CITATION = PHLTA,B295,371;%%
  %87 citations counted in INSPIRE as of 07 sept. 2015

%\cite{Blondel:2014bra}
\bibitem{Blondel:2014bra} 
  A.~Blondel {\it et al.} [FCC-ee study Team Collaboration],
  %``Search for Heavy Right Handed Neutrinos at the FCC-ee,''
  arXiv:1411.5230 [hep-ex].
  %%CITATION = ARXIV:1411.5230;%%
  %18 citations counted in INSPIRE as of 03 Oct 2015

%\cite{Abada:2014cca}
\bibitem{Abada:2014cca} 
  A.~Abada, V.~De Romeri, S.~Monteil, J.~Orloff and A.~M.~Teixeira,
  %``Indirect searches for sterile neutrinos at a high-luminosity
	%Z-factory,''
  JHEP {\bf 1504}  (2015) 051
  [arXiv:1412.6322 [hep-ph]].
  %%CITATION = ARXIV:1412.6322;%%
  %11 citations counted in INSPIRE as of 03 Oct 2015

%\cite{Abada:2013aba}
\bibitem{Abada:2013aba} 
  A.~Abada, A.~M.~Teixeira, A.~Vicente and C.~Weiland,
  %``Sterile neutrinos in leptonic and semileptonic decays,''
  JHEP {\bf 1402} (2014) 091 
  [arXiv:1311.2830 [hep-ph]].
  %%CITATION = ARXIV:1311.2830;%%
  %32 citations counted in INSPIRE as of 19 Jul 2015

%\cite{Abada:2012mc}
\bibitem{Abada:2012mc} 
  A.~Abada, D.~Das, A.~M.~Teixeira, A.~Vicente and C.~Weiland,
  %``Tree-level lepton universality violation in the presence of sterile
	%neutrinos: impact for $R_K$ and $R_\pi$,''
  JHEP {\bf 1302} (2013) 048 
  [arXiv:1211.3052 [hep-ph]].
  %%CITATION = ARXIV:1211.3052;%%
  %46 citations counted in INSPIRE as of 31 Aug 2015

%\cite{Asaka:2014kia}
\bibitem{Asaka:2014kia} 
  T.~Asaka, S.~Eijima and K.~Takeda,
  %``Lepton Universality in the $\nu$MSM,''
  Phys.\ Lett.\ B {\bf 742} (2015) 303 
  [arXiv:1410.0432 [hep-ph]].
  %%CITATION = ARXIV:1410.0432;%%
  %6 citations counted in INSPIRE as of 31 Aug 2015

%\cite{Antusch:2008tz}
\bibitem{Antusch:2008tz} 
  S.~Antusch, J.~P.~Baumann and E.~Fernandez-Martinez,
  %``Non-Standard Neutrino Interactions with Matter from Physics Beyond
	%the Standard Model,''
  Nucl.\ Phys.\ B {\bf 810} (2009) 369 
  [arXiv:0807.1003 [hep-ph]].
  %%CITATION = ARXIV:0807.1003;%%
  %144 citations counted in INSPIRE as of 19 Jul 2015

%\cite{Antusch:2014woa}
\bibitem{Antusch:2014woa} 
  S.~Antusch and O.~Fischer,
  %``Non-unitarity of the leptonic mixing matrix: Present bounds and
	%future sensitivities,''
  JHEP {\bf 1410} (2014) 94 
  [arXiv:1407.6607 [hep-ph]].

 \bibitem{Chanowitz:1978mv}
  M.~S.~Chanowitz, M.~A.~Furman and I.~Hinchliffe,
  %``Weak Interactions of Ultraheavy Fermions. 2.,''
  Nucl.\ Phys.\ B {\bf 153} (1979) 402.

\bibitem{Durand:1989zs}
  L.~Durand, J.~M.~Johnson and J.~L.~Lopez,
  %``Perturbative Unitarity Revisited: A New Upper Bound on the Higgs
  %Boson Mass,'' 
  Phys.\ Rev.\ Lett.\  {\bf 64} (1990) 1215.

\bibitem{Korner:1992an}
  J.~G.~Korner, A.~Pilaftsis and K.~Schilcher,
  %``Leptonic flavor changing Z0 decays in SU(2) x U(1) theories with
  %right-handed neutrinos,'' 
  Phys.\ Lett.\ B {\bf 300} (1993) 381
  [hep-ph/9301290].

\bibitem{Bernabeu:1993up}
  J.~Bernabeu, J.~G.~Korner, A.~Pilaftsis and K.~Schilcher,
  %``Universality breaking effects in leptonic Z decays,''
  Phys.\ Rev.\ Lett.\  {\bf 71} (1993) 2695
  [hep-ph/9307295].
    
\bibitem{Fajfer:1998px}
  S.~Fajfer and A.~Ilakovac,
  %``Lepton flavor violation in light hadron decays,''
  Phys.\ Rev.\ D {\bf 57} (1998) 4219.

\bibitem{Ilakovac:1999md}
  A.~Ilakovac,
  %``Lepton flavor violation in the standard model extended by heavy singlet Dirac neutrinos,''
  Phys.\ Rev.\ D {\bf 62} (2000) 036010
  [hep-ph/9910213].

  \bibitem{Blennow:2010th} 
  M.~Blennow, E.~Fernandez-Martinez, J.~Lopez-Pavon and J.~Menendez,
  %``Neutrinoless double beta decay in seesaw models,''
  JHEP {\bf 1007} (2010) 096 
  [arXiv:1005.3240 [hep-ph]].

  \bibitem{Abada:2014nwa}
  A.~Abada, V.~De Romeri and A.~M.~Teixeira,
  %``Effect of steriles states on lepton magnetic moments and
  %neutrinoless double beta decay,'' 
  JHEP {\bf 1409} (2014) 074
  [arXiv:1406.6978 [hep-ph]].
  %%CITATION = ARXIV:1406.6978;%%

 \bibitem{Agostini:2013mzu}
M.~Agostini {\it et al.}  [GERDA Collaboration],
%``Results on Neutrinoless Double-$\beta$ Decay of $^{76}$Ge from
%Phase I of the GERDA Experiment,'' 
Phys.\ Rev.\ Lett.\  {\bf 111} (2013) 12,  122503
[arXiv:1307.4720 [nucl-ex]].
%%CITATION = ARXIV:1307.4720;%%

\bibitem{Auger:2012ar} 
  M.~Auger {\it et al.}  [EXO Collaboration],
  %``Search for Neutrinoless Double-Beta Decay in $^{136}$Xe with EXO-200,''
  Phys.\ Rev.\ Lett.\  {\bf 109} (2012) 032505 
  [arXiv:1205.5608 [hep-ex]].
 
\bibitem{Albert:2014awa} 
  J.~B.~Albert {\it et al.}  [EXO-200 Collaboration],
  %``Search for Majorana neutrinos with the first two years of EXO-200 data,''
  Nature {\bf 510} (2014) 229-234 
  [arXiv:1402.6956 [nucl-ex]].
  
  \bibitem{Gando:2012zm} 
  A.~Gando {\it et al.}  [KamLAND-Zen Collaboration],
  %``Limit on Neutrinoless $\beta\beta$ Decay of Xe-136 from the First
  %Phase of KamLAND-Zen and Comparison with the Positive Claim in
  %Ge-76,'' 
  Phys.\ Rev.\ Lett.\  {\bf 110} (2013) 062502 
  [arXiv:1211.3863 [hep-ex]].

%\cite{Bennett:2006fi}
\bibitem{Bennett:2006fi} 
  G.~W.~Bennett {\it et al.} [Muon g-2 Collaboration],
  %``Final Report of the Muon E821 Anomalous Magnetic Moment Measurement
	%at BNL,''
  Phys.\ Rev.\ D {\bf 73} (2006) 072003 
  [hep-ex/0602035].
  %%CITATION = HEP-EX/0602035;%%
  %1019 citations counted in INSPIRE as of 25 Sep 2015
  


\end{thebibliography}
\end{document}